\newif\ifdraft
\drafttrue

\newif\ifarxiv
\arxivtrue 

\documentclass[journal,twoside,web]{ieeecolor}
\usepackage{generic}
\usepackage{cite}
\usepackage{amsmath,amssymb,amsfonts}
\usepackage{algorithmic}
\usepackage{graphicx}
\usepackage{textcomp}
\def\BibTeX{{\rm B\kern-.05em{\sc i\kern-.025em b}\kern-.08em
    T\kern-.1667em\lower.7ex\hbox{E}\kern-.125emX}}

\ifarxiv
\else 
\fi

\usepackage{hyperref}
\usepackage{bbm}
\usepackage{comment}
\usepackage{float}
\usepackage{amsfonts}
\usepackage{siunitx}
\newcommand\SSSDECG{\textit{SSSD-ECG}}

\usepackage[normalem]{ulem}
\newcommand{\stkout}[1]{\ifmmode\text{\sout{\ensuremath{#1}}}\else\sout{#1}\fi}

\ifdraft

\newcommand{\deleted}[1]{\textcolor{red}{\stkout{#1}}}

\newcommand{\deletedfloat}[1]{}
\newcommand{\commented}[1]{\textcolor{blue}{#1}}
\else

\newcommand{\deleted}[1]{}

\newcommand{\deletedfloat}[1]{}
\newcommand{\commented}[1]{}
\fi

\begin{document}
\title{Diffusion-based Conditional ECG Generation with Structured State Space Models}

\author{Juan Miguel Lopez Alcaraz and Nils Strodthoff
\thanks{\ifarxiv\else
Paper submitted for review on 28.03.2023. \\ \fi
Juan Miguel Lopez Alcaraz and Nils Strodthoff are with the University of Oldenburg, 26129 Oldenburg, Germany (e-mail: juan.lopez.alcaraz@uol.de, nils.strodthoff@uol.de). Corresponding author: Nils Strodthoff. }
}

\maketitle
\begin{abstract}
Generating synthetic data is a promising solution for addressing privacy concerns that arise when distributing sensitive health data. In recent years, diffusion models have become the new standard for generating various types of data, while structured state space models have emerged as a powerful approach for capturing long-term dependencies in time series. Our proposed solution, SSSD-ECG, combines these two technologies to generate synthetic 12-lead electrocardiograms (ECGs) based on over 70 ECG statements. As reliable baselines are lacking, we also propose conditional variants of two state-of-the-art unconditional generative models. We conducted a thorough evaluation of the quality of the generated samples by assessing pre-trained classifiers on the generated data and by measuring the performance of a classifier trained only on synthetic data. SSSD-ECG outperformed its GAN-based competitors. Our approach was further validated through experiments that included conditional class interpolation and a clinical Turing test, which demonstrated the high quality of SSSD-ECG samples across a wide range of conditions.
\end{abstract}


\section{Introduction}
\label{sec:introduction}


\IEEEPARstart{W}{ithout} a doubt, the vast amount of data generated worldwide has been a major catalyst for significant advancements in machine learning across a variety of data types and applications. However, acquiring and sharing data can present challenges, even among different departments within the same organization, due to privacy concerns. This is especially true in privacy-sensitive fields like healthcare.  Considerable efforts have been made to enhance the security and privacy of healthcare data through the implementation of regulations such as GDPR or HIPAA \cite{gdpr, hippa}, compliance measures \cite{ecloud, ehealthrecordsystem, bigdatahealthcare}, as well as the development of technical solutions such as blockchain technology or federated learning \cite{blockchainhealth, federatedlearning}. However, even technical solutions like federated learning alone do not guarantee complete privacy protection, as trained models can be reconstructed to reveal training samples through model inversion attacks \cite{yin2021see}. Therefore, a combination of different privacy-enhancing techniques must be employed.

The ability to create digital replicas of raw data, known as digital twins, is a viable solution that preserves the statistical properties of the original data while removing personal patient information. Data augmentation techniques, such as statistical methods, provide limited privacy protection by disguising the original data. Therefore, the use of generative machine learning models has become increasingly popular. However, it is important to ensure that the recreated data is accurate, otherwise it may be biased \cite{MADLEYDOWD201963} and negatively impact downstream tasks and decision-making, including interpretability \cite{classificationwithimputedmissingvalues}. This highlights the need for generative models that can produce high-quality samples based on different patient characteristics, as well as the development of objective benchmarking criteria to evaluate the quality of the generated samples.

Recently, diffusion models have exhibited remarkable outcomes for data synthesis, surpassing other models such as generative adversarial networks (GANs) or autoregressive models. These models have been shown to have biases towards high-density classes \cite{ganslowdensity}, as well as low fidelity dataset distributions \cite{NEURIPS2021_49ad23d1, autoregressivelowfidelity}, and are known to suffer from training instabilities \cite{ecg_from_gans}. On top of that, diffusion models have shown improvements on various downstream tasks \cite{conditionalgenerativeaudiodatasets}, and have assisted in the identification of new pathologies \cite{deepfakedigitalpatology}.

In this study, our focus is on generating synthetic electrocardiogram data. The ECG is a commonly used medical procedure due to its non-invasive nature, simple and reliable technology, and high diagnostic value. It is an essential tool for the initial assessment of a patient's overall cardiac state. This importance is further highlighted by recent advances in AI-based ECG analysis, as discussed in \cite{TOPOL2021785}. However, many high-impact studies on ECG analysis have been conducted using private datasets, which poses a significant problem in terms of reproducibility and hinders progress in the research community as a whole. To enable the sharing of such datasets, high-quality digital twins of private datasets, which contain highly sensitive patient data and are subject to strict privacy requirements, need to be generated. We demonstrate this process by creating and evaluating synthetic copies of the publicly accessible PTB-XL dataset \cite{Wagner:2020PTBXL,Wagner2020:ptbxlphysionet,Goldberger2020:physionet}, which is a popular ECG dataset.

The main contributions of this work are the following:(1) We propose a diffusion model for generating short (10s) 12-lead ECGs. This model uses a structured state space model as its internal component. (2) We introduce conditional variants of two state-of-the-art unconditional generative models for ECG data. (3) We generate synthetic versions of PTB-XL, a large publicly available ECG dataset, and evaluate the quality of the generated samples by training and testing classifiers on these datasets. (4) We demonstrate that our model has internalized domain knowledge in several ways: a) by comparing generated samples using a beat-level aggregation across subgroups of samples with common pathologies, b) by meaningfully interpolating between different sets of conditions, and c) by conducting an expert assessment in the form of a clinical Turing test that confirms the high quality of the generated samples.

\section{Materials \& Methods}
\subsection{Dataset and downstream task} 

The focus of this study is on short 12-lead ECGs that last 10 seconds. These ECGs are obtained from six limb leads and six precordial leads, which is the most commonly used method in clinical practice. The ECGs were sampled at a rate of 100~Hz, as previous research has shown that increasing the sampling rate to 500~Hz does not significantly improve the ability to classify ECGs \cite{Mehari:2022S4}.

\subsubsection*{PTB-XL dataset}
Our experiments are based on the PTB-XL dataset \cite{Wagner:2020PTBXL,Wagner2020:ptbxlphysionet,Goldberger2020:physionet}, which is a publicly available collection of clinical 12-lead ECG data comprising 21,837 records from 18,885 patients. In order to train a high-quality generative model, it is important to have a dataset of sufficient size. For class-conditional generative models, which require sample-wise annotations for all samples in the dataset similar to supervised discriminative training, PTB-XL is a good choice as it provides annotations for each sample in terms of 71 ECG statements in a multi-label setting. These cover 44 diagnostic (organized into 24 sub-classes and 5 superclasses comprising normal, conduction disturbance, myocardial infarction, hypertrophy, ST/T changes), 19 form-related (5 of which also counted as diagnostic statements), such as abnormal QRS-complex, and 12 rhythm-related  statements, such as atrial fibrillation. It is worth noting that this dataset represents a significant advancement in terms of complexity, as it includes a broad set of 71 ECG statements, which can be used in a multi-label setting. Most literature approaches have typically focused on a single condition, such as healthy samples, or on generating samples based on very limited sets of class labels. For further information about the dataset, please refer to Appendix~\ref{app:ptbxl} and the dataset descriptor \cite{Wagner:2020PTBXL}.

\subsubsection*{Downstream task}
As downstream task, we are examining the problem of predicting ECG statements at the most granular level, which involves classifying them into multiple labels. This task is a well-researched benchmark on the PTB-XL dataset, and we follow the established methodology outlined in previous work \cite{deep_ptbxl}. In particular, we make use of the proposed stratified folds for model training, where the first 8 folds are used for training, the 9th fold serves as validation set and the 10th fold as test set. Our evaluation metric for this task is the macro-averaged area under the receiver operating curve (macro AUROC) across all 71 labels on the PTB-XL test set. As model architecture, we use a XResNet1d50 model, which is a one-dimensional adaptation of a modern ResNet model \cite{he2019bag}, as proposed in \cite{deep_ptbxl}. We closely follow the training and evaluation methodology outlined in \cite{deep_ptbxl}, with a binary crossentropy loss as our training objective, which is appropriate for a multi-label classification problem. We train the model on random crops of 2.5 seconds in length and during test time, we average seven overlapping predictions from the same sample to obtain the final prediction for the sample. To prevent overfitting, irrespective of whether we train on real or synthetic data, we always perform model selection based on a corresponding real/synthetic validation set score (macro AUC). For further information on the downstream classifier and training, please refer to Appendix~\ref{app:xresnet}.

\subsection{Background}
\label{sec:background}

\subsubsection*{Diffusion models} 
Diffusion models \cite{pmlr-v37-sohl-dickstein15}, which are a type of generative model, have shown state-of-the-art performance on a variety of data modalities, including audio data \cite{chen2020wavegrad,DBLP:conf/iclr/KongPHZC21}, image data \cite{NEURIPS2021_49ad23d1,NEURIPS2020_4c5bcfec,Ho2022CascadedDM,rombach2022high}, and video data \cite{ho2022video}. Diffusion models consist of two processes: the forward process and the backward process. During the forward process, noise is incrementally introduced in a Markovian manner. In contrast, during the backward process, the model gradually removes the noise. The forward process is parameterized as

\begin{equation}
\label{eq: forward}
q(x_{1}, \ldots , x_{T} | x_{0}) = \prod_{t=1}^{T} q(x_{t} | x_{t-1})\,,
\end{equation}

where $q(x_{t}|x_{t-1})=\mathcal{N}(x_t;\sqrt{1-\beta_t} x_{t-1},\beta_t \mathbbm{1})$, $\beta_t$ are (fixed or learnable) forward-process variances, which adjust the noise level and $T$ is the number of diffusion steps. Equivalently, $x_t$ can be expressed in closed form as $x_t = \sqrt{\alpha_t} x_0 + (1-\alpha_t)\epsilon$ for $\epsilon\sim \mathcal{N}(0,\mathbbm{1})$, where $\alpha_t = \sum_{i=1}^t (1-\beta_t)$. 
The backward process is parameterized as in Equation~\eqref{eq: backward}, where $x_T\sim \mathcal{N}(0,\mathbbm{1}$). 

\begin{equation}
\label{eq: backward}
p_\theta(x_{0}, \ldots, x_{t-1} | x_{T}) = p(x_T)  \prod_{t=1}^{T} p_\theta(x_{t-1} | x_t)       
\end{equation}

Using a particular parameterization of $p_\theta(x_{t-1} | x_t)$, it was shown in \cite{NEURIPS2020_4c5bcfec} that the reverse process can be trained using
\begin{align}
\label{eq:objective}
L= \text{min}_\theta & \mathbbm{E}_{x_0\sim \mathcal{D},\epsilon\sim\mathcal{N}(0,\mathbbm{1}),t\sim \mathcal{U}(1,T)} \\
&\quad\quad||\epsilon - \epsilon_\theta(\sqrt{\alpha_t}x_0+(1-\alpha_t)\epsilon,t)||_2^2 \nonumber\,,
\end{align}

where $\epsilon_\theta(x_t,t)$ is parameterized using a neural network and $\mathcal{D}$ denotes the data distribution.
This objective can be understood as a weighted variational bound on the negative log-likelihood that reduces the significance of terms at low $t$, i.e., at low noise levels. Class-conditional diffusion models can be realized by conditioning the backward process on desired set of labels $c$, i.e., using $\epsilon_\theta=\epsilon_\theta(x_t,t,c)$.

\subsubsection*{Structured state space models}
In essence, structured state space models (SSSMs) rely on a linear state space transition equation that links a one-dimensional input sequence, denoted as $u(t)$, with a one-dimensional output sequence, denoted as $y(t)$, by way of a hidden state $x(t)$ that is of $N$ dimensions,

\begin{align}
\label{eq: ssms}
x'(t) &= A x(t) + B u(t)  \\  
y(t) &= C x(t) + D u(t)  \nonumber\,,
\end{align}

where $A,B,C,D$ are transition matrices. In \cite{Gu2021EfficientlyML}, it is discussed that the SSSM (Structured State Space Model) is a promising model for capturing long-term dependencies in time series. This is achieved by discretizing the input and output relations and representing them as a convolution operation. This operation can be effectively computed on modern GPUs using a custom kernel. The ability to capture long-term dependencies is closely linked to the initialization of the hidden-to-hidden transition matrix $A$, as discussed in \cite{NEURIPS2020_102f0bb6}.By stacking multiple SSSM blocks, each with appropriate normalization and point-wise fully-connected layers, in a manner similar to a transformer layer, one can create a Structured State Space Sequence Model (S4). The S4 model has demonstrated outstanding performance on a variety of long-range-interaction benchmarks and sequence classification tasks, including 12-lead ECG classification, as shown in the most recent work \cite{Mehari:2022S4}.

\subsubsection*{Related work}
Deep generative modeling for time series data is an emerging subfield in machine learning, driven mainly by the constant progress in the development of generative models, particularly in the field of imaging. While various algorithm architectures have been developed, such as variational autoencoders \cite{9761431} and stacked denoising autoencoders \cite{RAHHAL2016340}, generative adversarial networks (GANs) remain the dominant technology in the field, especially with recurrent neural networks (RNNs) \cite{deepfakeelectrocardiograms,Zhu2019ElectrocardiogramGW,https://doi.org/10.48550/arxiv.1909.09150,Golany_Lavee_Tejman_Yarden_Radinsky_2020}, transformers \cite{10.1007/978-3-031-09342-5_13}, or differential equations \cite{pmlr-v119-golany20a} as building blocks. Although some prior research has addressed conditional time-series generation \cite{9761431, Zhu2019ElectrocardiogramGW, pmlr-v119-golany20a, Golany_Radinsky_2019, Golany_Lavee_Tejman_Yarden_Radinsky_2020}, most of it is still limited to the unconditional setting \cite{NEURIPS2019_c9efe5f2, 10.1007/978-3-031-09342-5_13, deepfakeelectrocardiograms, https://doi.org/10.48550/arxiv.1909.09150, adib2023synthetic}.

During the final stage of the manuscript preparation, we became aware of \cite{adib2023synthetic}, whose authors also applied a diffusion model for synthetic ECG generation, albeit using image representations rather than the time series directly and for the restricted case of the unconditional generation of single-channel ECGs. Interestingly, this approach was unable to outperform a GAN-based baseline according to several quality metrics. Also very recently, Chung et al proposed a  conditional generative model to generate synthetic ECGs from text reports \cite{chung2023texttoecg}. Due to the different task setup, their results are not directly comparable to ours but represent an interesting direction for future research. Nevertheless, we see it as an import next step to establish proper baselines for the more controlled case of ECG generation based on structured labels first.

Apart from these very recent works, numerous literature approaches have attempted to address the issue of creating synthetic ECG signals, but these approaches are often subject to significant limitations. Firstly, many of these models are only able to generate short time series \cite{Golany_Lavee_Tejman_Yarden_Radinsky_2020, https://doi.org/10.48550/arxiv.1909.09150, NEURIPS2019_c9efe5f2, 10.1007/978-3-031-09342-5_13}, which is why many of them restrict to the generation of single beats. Secondly, they are frequently trained on small sets of data from few patients \cite{Zhu2019ElectrocardiogramGW} or with limited conditional labels \cite{9761431,pmlr-v119-golany20a,Golany_Radinsky_2019}. Thirdly, many of these approaches require ECG beat segmentation as a preprocessing step, rather than working directly on continuous signals \cite{10.1007/978-3-031-09342-5_13, 9761431, pmlr-v119-golany20a,Golany_Lavee_Tejman_Yarden_Radinsky_2020}. Moreover, many of these approaches suffer from different evaluation issues, such as being limited to patient-specific generation and classification \cite{Golany_Radinsky_2019} or lacking training data and software for filtering/feature extraction for the general public \cite{deepfakeelectrocardiograms}. 

\subsection{Conditional generative models for ECG data}

\begin{figure}[ht]
\centerline{\includegraphics[width=\columnwidth]{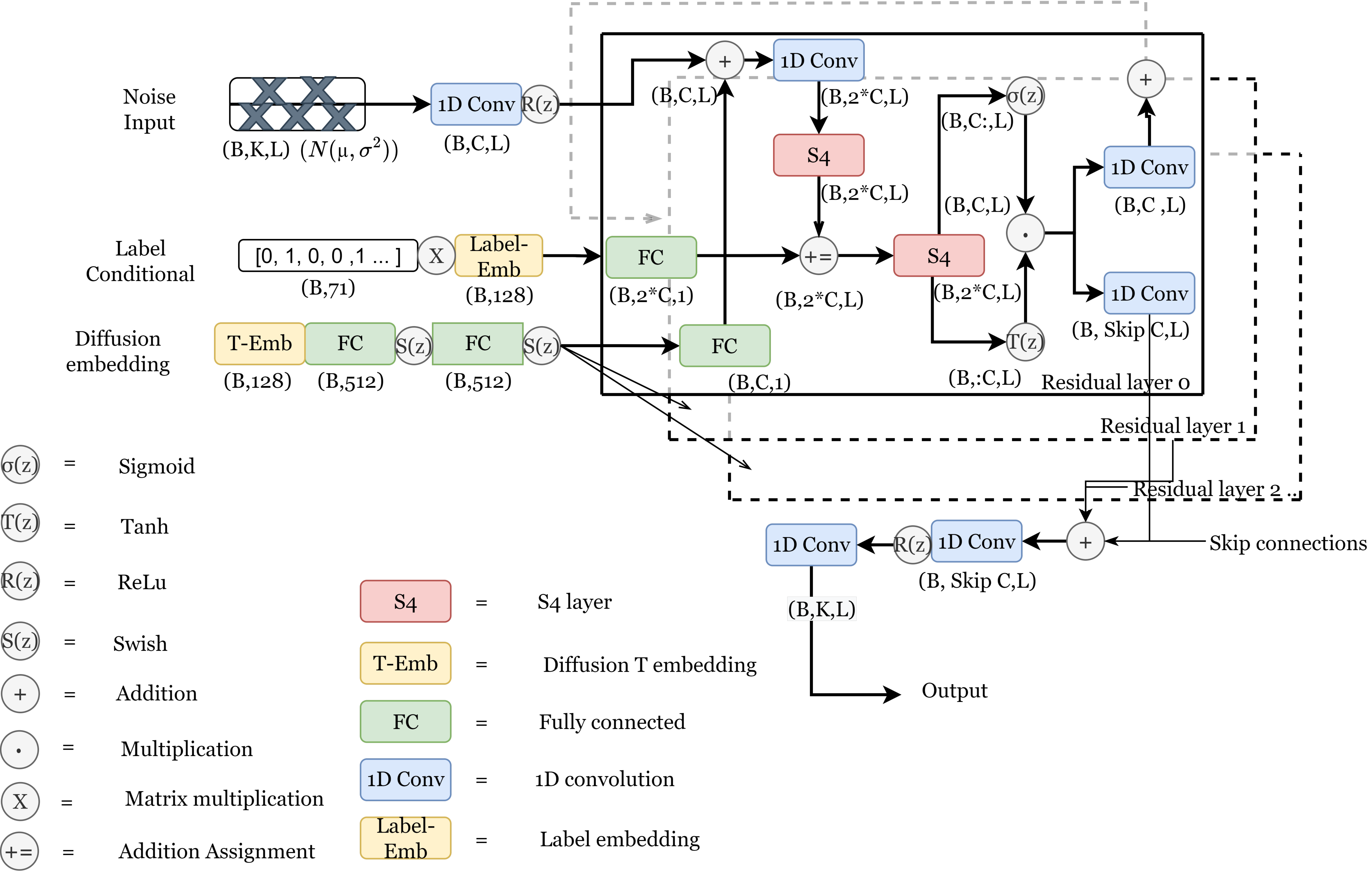}} 
\caption{Schematic representation of the \SSSDECG{} model architecture.}
\label{architecture}
\end{figure}

\subsubsection*{\SSSDECG{}}

Returning to the discussion of probabilistic diffusion in Section~\ref{sec:background}, the only remaining component that needs to be specified is the explicit parameterization of the backward process, denoted by $\epsilon_\theta=\epsilon_\theta(x_t,t,c)$. In this regard, we have implemented an adaptation of the recently proposed $SSSD^{S4}$ model \cite{lopezalcaraz2022diffusionbased}, where SSSD stands for structured state space diffusion. The model is built on the DiffWave architecture \cite{DBLP:conf/iclr/KongPHZC21}, which was originally proposed for audio synthesis using dilated convolutions. However, in $SSSD^{S4}$, dilated convolutions are replaced by two S4 layers (as described in Section~\ref{sec:background}) to handle long-term dependencies better in time series. This has been shown to be effective for various time series imputation across different scenarios and forecasting tasks in \cite{lopezalcaraz2022diffusionbased}. \SSSDECG{}, which is proposed in this work, builds on the $SSSD^{S4}$ model architecture but also deviates from it in several important aspects. The most important difference is due to the different conditional information that is provided for respective applications and the omission of task specific masking strategies. While $SSSD^{S4}$ receives an imputation mask and the remaining input signal as conditional information, \SSSDECG{} is conditioned on the set of annotated ECG statements, which is in our case encoded as a binary vector of length 71. This vector is transformed into a continuous representation by multiplying it with a learnable weight matrix, transformed through a fully connected layer and is subsequently passed as conditional information to different $SSSD^{S4}$ layers. For more details about the model's internals and hyperparameters, please refer to Appendix~\ref{app:sssdecg}.

As a further task-specific modification, it is worth mentioning that in a standard 12-lead ECG, only two out of the six limb leads are independent. This means that any set of limb leads can be reconstructed using any two given limb leads, based on the defining relationships III = II-I, aVL = (I-III)/2, aVF = (II+III)/2, and -aVR = (I+II)/2. To ensure that the ECGs we generate satisfy these relationships, we use generative models to synthesize only 8 leads - the 6 precordial leads and 2 limb leads (I and aVF in our case). We then reconstruct the remaining 4 leads by sampling from the two limb leads (I and aVF). This approach is similar to the one used by \cite{deepfakeelectrocardiograms} and is also applied to all baseline models described below.

We believe that using a publicly available ECG datasets for training is a critical step toward measurable progress. This allows for improvements in model architecture and training schedules to be disentangled from improvements resulting solely from larger or more comprehensive training datasets. In contrast to previous limitations, the \SSSDECG{} model generates long sequences of 1000 time steps (for a 10-second ECG at 100 Hz), is trained on a large dataset of over 18,000 patients, and is conditioned on a rich set of 71 ECG statements. Additionally, it generates full samples without the need for prior segmentation or any other preprocessing and builds on publicly available code \cite{coderepo}.

\subsubsection*{Baselines: WaveGAN* and Pulse2Pulse}

Our primary contribution is the \SSSDECG{} model, but we also present conditional versions of two existing generative models for ECG data, namely WaveGAN and Pulse2Pulse \cite{deepfakeelectrocardiograms}. These models use Generative Adversarial Networks (GANs) \cite{goodfellow2020generative}. We make these models class-conditional by incorporating batch normalization layers into their architecture and converting them into conditional batch normalization layers \cite{de2017modulating}. This involves making the layer's internal shift and scaling parameters dependent on the class. We follow a similar approach to map the binary label vector into a continuous representation using a learned weight matrix. Further details about the model configurations and training hyperparameters are provided in Appendix~\ref{app:baselines}.

Regrettably, we were unsuccessful in training effective generative models for our particular case using publicly available implementations, except for WaveGAN and Pulse2Pulse as mentioned earlier. Our attempts included creating generative models for time series generation, such as TTS-GAN \cite{10.1007/978-3-031-09342-5_13} and well-established methods like TimeGAN \cite{NEURIPS2019_c9efe5f2}, but these models were likely challenged by input lengths of 1000 time steps. We were also unable to generate samples at 250 time steps. Additionally, we were not successful in training a class-conditional model using the cVAE\_ECG \cite{9761431} approach, as described in Appendix~\ref{app:unrelieable_baselines}.

\subsection{Performance measures for generative models}
\label{sec:metrics}

In this section, our objective is to propose measures to quantify the quality of the generated samples.
We can achieve this by training classifiers on either real or synthetic training sets and testing them on either real or synthetic test sets. It is worth noting that we only need to train a single classifier on real data, which we will refer to as the reference classifier, which is kept fixed for all the remaining experiments. From it, one can infer three different performance measures:

\begin{enumerate}
\item The primary criterion for evaluating the quality of synthetic data is its \textit{capacity to replace real data}. This evaluation involves testing a classifier that is trained on a synthetic training set with a real test set. The ranking of various algorithms based on this measure is widely regarded as the gold standard for assessing the quality of synthetic data models, as mentioned in \cite{pmlr-v162-alaa22a}. However, it is also valuable to compare the absolute performance of the synthetic classifier with that of the reference classifier trained on real data.

\item The second metric, which complements the first, aims to determine \textit{the realism of the synthetic data by evaluating it using a reference classifier}. This involves using the reference model, which was trained on real data, to assess the performance of the synthetic test set. The expected decrease in predictive performance, as compared to the evaluation using the real test set, is due to the inherent difference between the distribution of the real training data and that of the synthetic test data. This decrease can serve as a second measure of the performance of generative models.

\item The third performance indicator takes a distinct approach by evaluating the \textit{internal consistency}. It assesses how well a classifier trained on synthetic training data generalizes to unseen synthetic test data from the same distribution. To evaluate this, a classifier is trained on a synthetic training set and tested on its corresponding synthetic test set. However, this criterion does not reference the real data directly or indirectly through a reference classifier trained on real data, which is why it is considered less informative than the first two criteria.

\end{enumerate}

We would like to draw attention to recent research such as \cite{pmlr-v162-alaa22a} that focuses on evaluating the performance of generative models, specifically in producing output that matches a predetermined standard. However, during our attempts to replicate the findings, we encountered problems with instability when training one-class embedding as a necessary step in \cite{pmlr-v162-alaa22a}, which proved to significantly affect the results. Therefore, we have chosen not to include those corresponding results in our paper.

\section{Experiments and Results}
We trained three generative models - \SSSDECG{}, WaveGAN*, and Pulse2Pulse - based on the eight PTB-XL training folds. Each model was conditioned on the respective sample annotations in the dataset. After training each model, we generated a synthetic copy of the PTB-XL dataset as follows: For each sample in PTB-XL, we generated a new synthetic sample by conditioning on the ECG annotation of that sample. For each of the three generative models, this resulted in a synthetic ECG dataset that matches PTB-XL in terms of size and label distribution. In this way, we obtain synthetic training, validation and test sets, whose label distributions match exactly the corresponding PTB-XL sets. We used these synthetic datasets for both qualitative and quantitative assessments of the quality of the generated samples.

\subsection{Qualitative assessment}
To conduct a qualitative assessment, we chose two commonly occurring conditions, namely normal ECGs (NORM) and ECGs with left-ventricular hypertrophy (LVH) In order to be able to compare across samples, we perform beat-level segmentation across 250 generated samples for each condition by identifying R-peaks and cropping from 300ms before the R-peak until 500ms after the R-peak. Now, we plot the median across all beats along with the corresponding 0.25 and 0.75 quantiles visualized through a shaded band. This allows for a direct visual comparison of the main characteristics of the generated samples, both across different models and in comparison to the real samples from PTB-XL.

\begin{center}
\begin{figure}[!ht]
\includegraphics[width=\columnwidth]{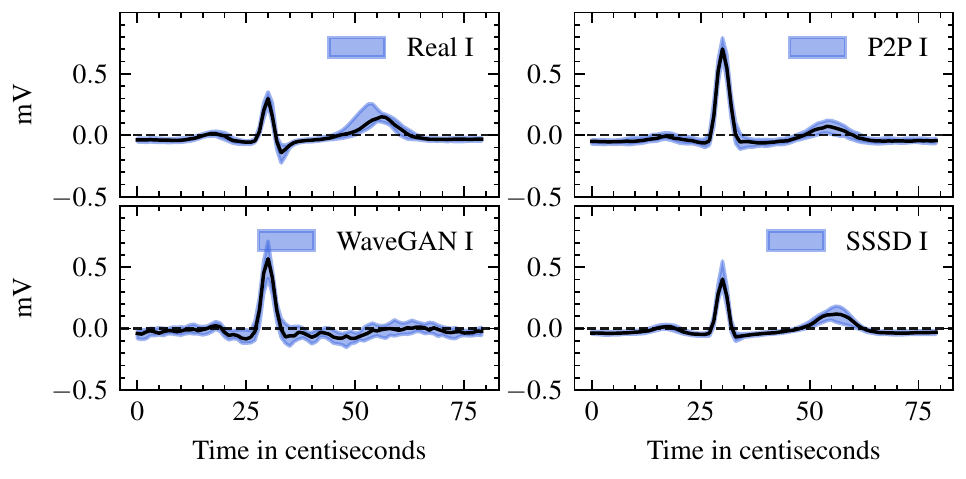}    
\includegraphics[width=\columnwidth]{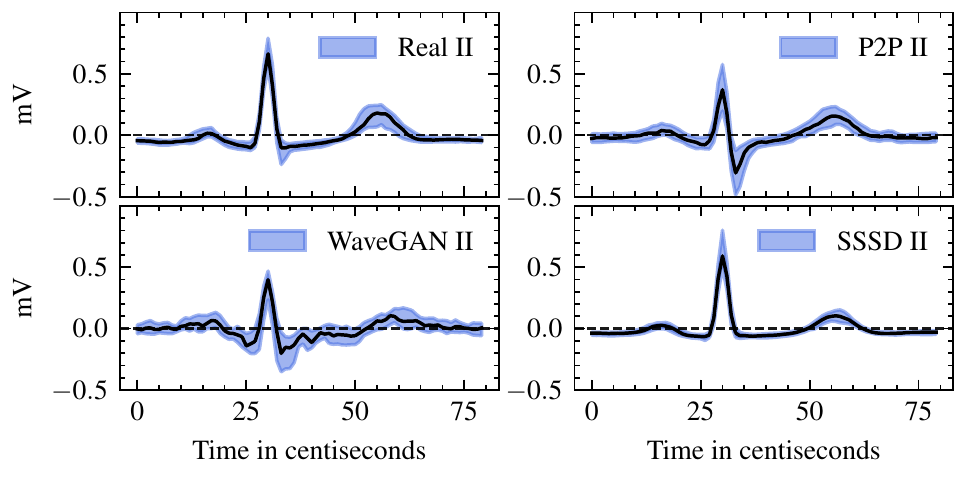}   
\includegraphics[width=\columnwidth]{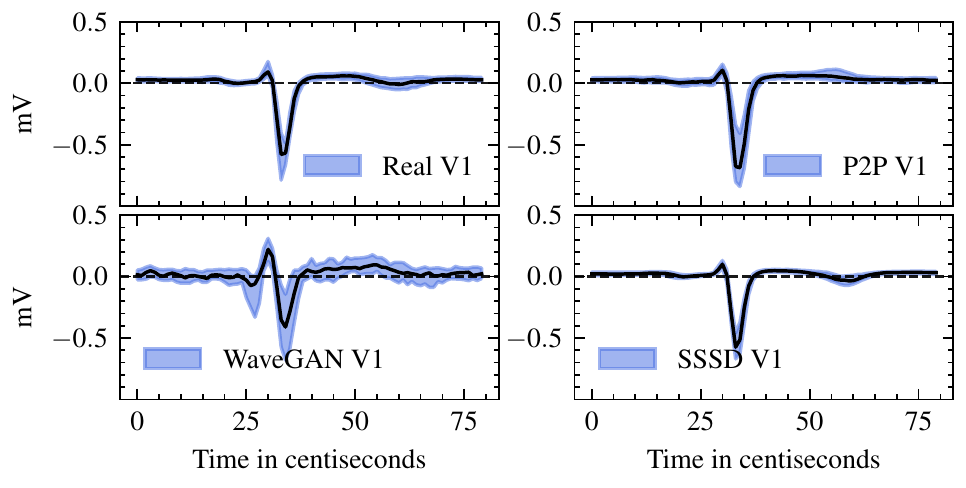}  
\includegraphics[width=\columnwidth]{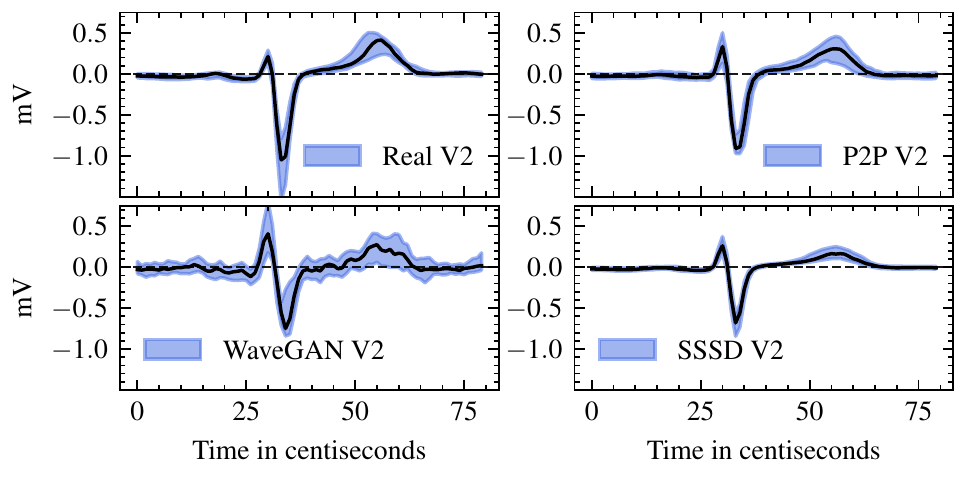} 
\caption{From top to bottom four grids of leads I, II, V1, and V2 respectively, where for each grid from top to bottom and left to right real, WaveGAN*, Pulse2Pulse, and \SSSDECG{} Median (black line) and 0.25-0.75 quantiles (blue shaded area) segmented beats for healthy (NORM) samples.}
\label{fig:qualitative_assessment_norm}
\end{figure}
\end{center}

Figure~\ref{fig:qualitative_assessment_norm} presents a qualitative evaluation of healthy (NORM) generated samples compared to real samples. The WaveGAN* model produces signals with a significant amount of interference and numerous ECG features that differ from those found in real signals, such as larger R-peaks and absent P- and T-waves in most leads due to signal inference. The Pulse2Pulse model produces more consistent results, with less variability across quantiles. However, there are some mismatched features compared to real samples, including the absence of an S-peak in lead I, a larger S-peak in lead II, and an opposite (upward) T-wave in lead V1. The \SSSDECG{} model is the one that closely resembles real samples, with a higher level of confidence as the quantile bands closely approximate the median in all features. This model correctly replicates the P- and T-waves in leads I and II, as well as the downward T-wave in lead V1. In addition, the features in lead V2, such as the R- and S-peaks and P- and T-waves, are balanced.

\begin{center}
\begin{figure}[!ht]
\includegraphics[width=\columnwidth]{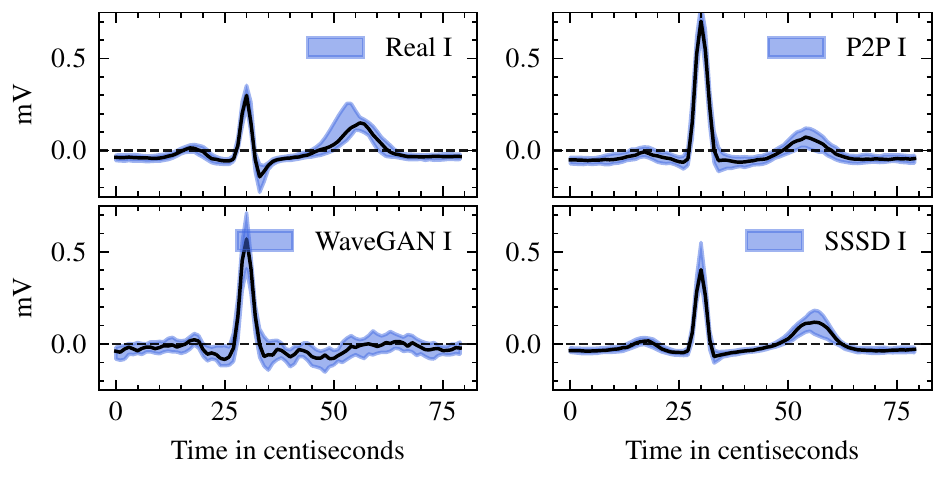} 
\includegraphics[width=\columnwidth]{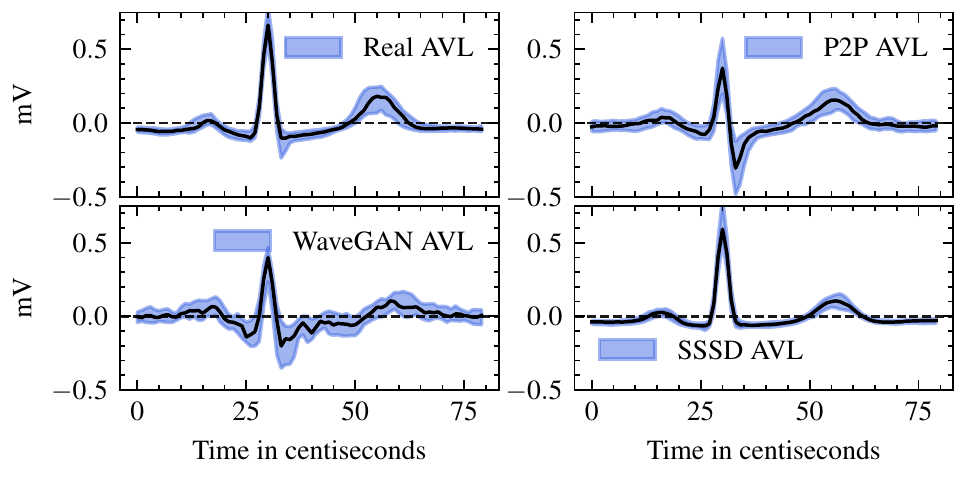}
\includegraphics[width=\columnwidth]{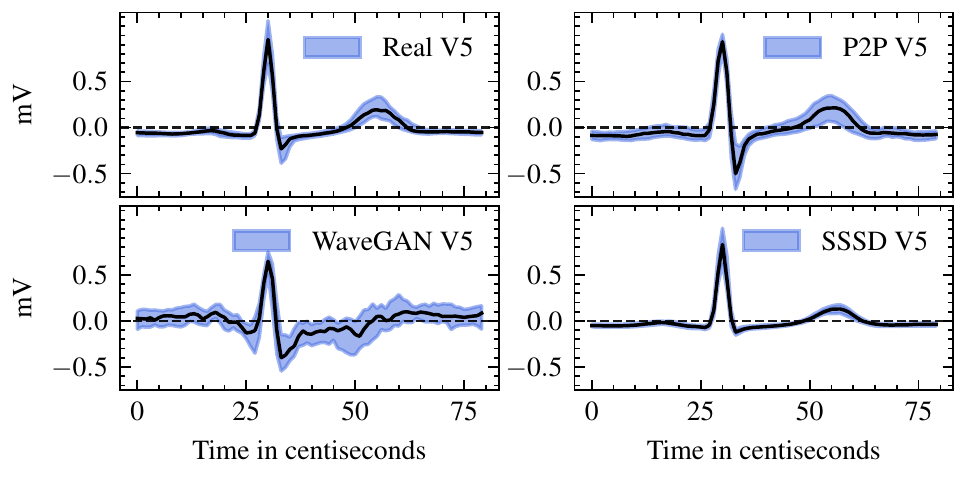} 
\includegraphics[width=\columnwidth]{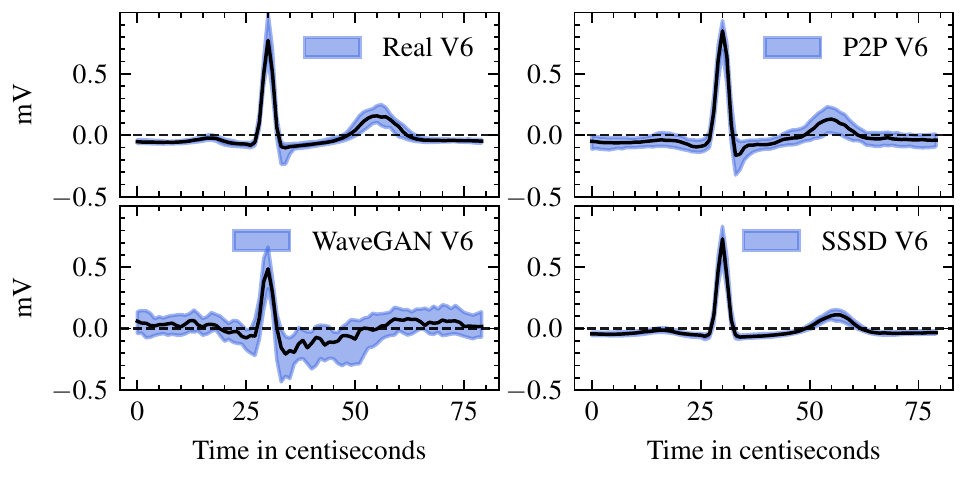} 
\caption{From top to bottom four grids of leads I, AVL, V5, and V6 respectively, where for each grid from top to bottom and left to right real, WaveGAN*, Pulse2Pulse, and \SSSDECG{} Median (black line) and 0.25-0.75 quantiles (blue shaded area) segmented beats for left-ventricular hypertrophy (LVH) samples.}
\label{fig:qualitative_assessment_lvh}
\end{figure}
\end{center}

Figure~\ref{fig:qualitative_assessment_lvh} shows a comparison of synthetic samples with left-ventricular hypertrophy (LVH) generated by different proposed models to real samples, based on a qualitative assessment. The WaveGAN* model has a lot of signal interference, which makes it difficult to observe certain features such as P- and T-waves, although the R-peak seems to be correctly shaped. In contrast, the Pulse2Pulse model fails to accurately generate the R-peak, which appears curved towards the top in all leads, and also fails to generate many other features, such as a straight T-wave in V5 and a downward T-wave in V6, with a high degree of uncertainty. Finally, the \SSSDECG{} plots are the most closely aligned with the real samples, generating a correct R-peak with relatively short P-waves and balanced T-waves for all beats, with consistent confidence intervals across all leads.

In summary, this initial qualitative evaluation provides indications of the superiority of the \SSSDECG{} samples over its competitors and is consistent with the characteristics of the corresponding real samples from PTB-XL.

\subsection{Quantitative assessment}

\begin{table}[ht]
    \caption{Classification performance of XResNet50 models trained/evaluated on different combinations of real/synthetic data.}
    \centering
    \setlength{\tabcolsep}{3pt}
    \begin{tabular}{|p{50pt}|p{50pt}|p{50pt}|}
    \hline
    Model                   & \multicolumn{2}{c|}{AUROC}  \\ \hline \hline
    
    WaveGAN*                &  Test real &  Test synth.        \\ \hline
    Train real      &       0.9317       &   0.6489              \\ 
    Train synth.     &     0.5816      &   0.9793           \\ \hline\hline

    Pulse2Pulse             &  Test real &  Test synth.        \\ \hline
    Train real      &       0.9317       &    0.7082         \\ 
    Train synth.      &       0.5968        &  \textbf{0.9950}                \\ \hline\hline

    \SSSDECG{}              &  Test real &  Test synth.        \\ \hline
    Train real      &        0.9317     &         \textbf{0.9434}          \\ 
    Train synth.      &        \textbf{0.8402}      &       0.9822          \\ \hline
\end{tabular}
\label{tab: aurocs}
\end{table}

In Table~\ref{tab: aurocs}, we present a quantitative comparison of the three proposed generative models based on the performance metrics introduced in Section~\ref{sec:metrics}. The first metric shows that \SSSDECG{} outperforms its competitors with a score of 0.84, compared to only 0.60 for Pulse2Pulse and 0.58 for WaveGAN*. One possible explanation for this result is that Pulse2Pulse and WaveGAN* are both GAN-based approaches, which tend to focus on selected nodes rather than covering the full distribution. The second metric also shows a similar pattern, with \SSSDECG{} clearly outperforming Pulse2Pulse (0.71) and WaveGAN* (0.64) with a score of 0.94. According to the third metric, all three approaches achieve scores of 0.98 or higher, indicating a high degree of internal consistency among the models. In summary, the results of our experiment demonstrate a clear quantitative advantage of \SSSDECG{} over its GAN-based competitors. We will now reflect in more detail on the quantitative performance of \SSSDECG{} and its implications:

Remarkably, on one hand, \SSSDECG{} even achieves a slighter better score (0.94) assessed through the reference classifier compared to evaluating the reference classifier on real samples. This is an encouraging prospect for auditing ECG analysis algorithms, as pre-screening with synthetic data may be conducted before evaluating them on high-quality private test data.

On the flip side, it is important to acknowledge that the \SSSDECG{} performance during training with synthetic data and testing with real data is notably lower compared to the reference classifier trained on actual data (0.84 vs. 0.93). It is important to reiterate the difficulty of the generation task at hand. This work is the first attempt at constructing a generative model that is dependent on a diverse range of 71 ECG statements. Additionally, one must not underestimate the fact that these 71 statements are employed in a multi-label environment, where the number of unique label combinations (including co-occurring conditions) is significantly higher, which is a reflection of the complex reality of co-existing diseases and disease states. Some of the ECG statements are already sparsely populated with less than 100 occurrences throughout the entire dataset, and this is even more pronounced in the case of co-occurring label combinations. We consider it to be a challenging but worthwhile objective for the research community to devise methods that will bridge the gap in the ``train on synthetic, test on real'' situation, ultimately allowing synthetic data to be used essentially interchangeably with real data.

\subsection{Conditional class interpolation}
In order to show that the \SSSDECG{} model has gained valuable knowledge in the specific field of ECG analysis, we conducted various class interpolation experiments. Unlike other models, our model is not limited to using only binary vectors as conditional information. Instead, we can use any real-valued vectors with values ranging from 0 to 1. By using a convex combination of two binary annotation vectors A and B, specified by $\alpha a+(1-\alpha) b$ for $\alpha\in[0,1]$, we can interpolate between the two conditions. The parameter $\alpha$ determines the weight given to condition A. The sample's initialization is kept constant throughout the process. By varying $\alpha$, we can smoothly transition between the two conditions. To better illustrate this, we divided the generated samples based on R-peaks and only presented median beats extracted from the signal. This makes it easier to observe signal changes as we move from one condition to another. This not only serves as an interesting consistency check, but it also opens up possibilities for more complex generative models that can incorporate non-binary disease states.

\subsubsection*{Inferior myocardial infarction}
\begin{figure}[!ht]
\centerline{\includegraphics[width=\columnwidth]{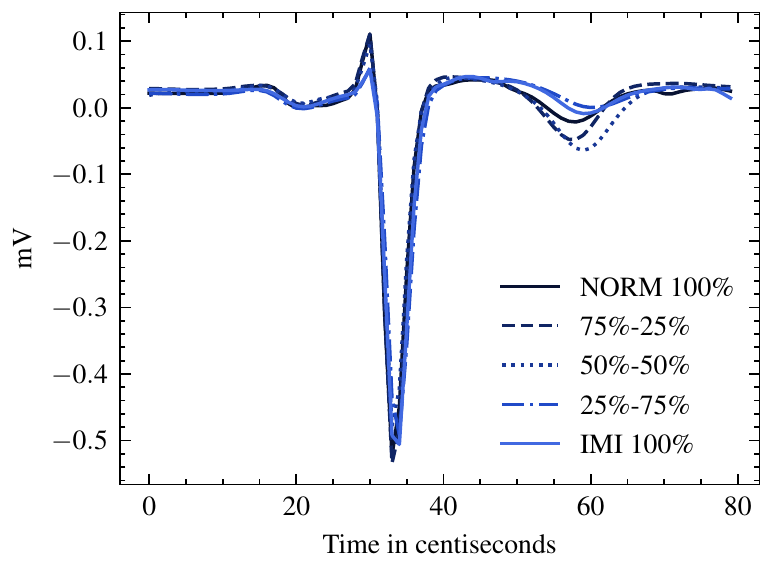}} 
\caption{\SSSDECG{} interpolation between a healthy (NORM+SR) and an inferior myocardial infarction (IMI+SR+ABQRS) signal in lead V1, where 5 signals of 100\% and 0\%, 75\%, and 25\%, 50\% and 50\%, 25\% and 75\%, and 0\% and 100\% of healthy and inferior myocardial infarction, respectively.}
\label{fig1}
\end{figure}

Figure~\ref{fig1} shows the interpolation between a healthy normal sample (NORM) and a signal from an inferior myocardial infarction (IMI) for lead V1. The labels for the healthy samples are based on their occurrence in the training sample and represent NORM and sinus rhythm (SR). The labels for the IMI disease include SR, IMI, and abnormal QRS complex (ABQRS). It can be observed that the generated IMI signals and their interpolations exhibit an early Q-wave formation, whereas the normal signal has a shorter and downward shape.

\subsubsection*{Complete left bundle branch block}
\begin{figure}[!ht]
\centerline{\includegraphics[width=\columnwidth]{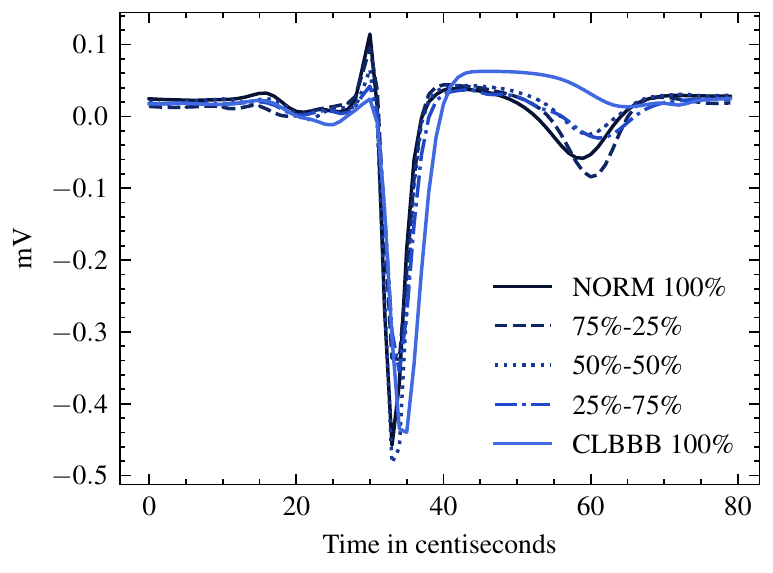}} 
\caption{\SSSDECG{} interpolation between a healthy (NORM+SR) and a complete left bundle branch block signal (CLBBB+SR) in lead V1, where 5 signals of 100\% and 0\%, 75\%, and 25\%, 50\% and 50\%, 25\% and 75\%, and 0\% and 100\% of healthy and complete left bundle branch block, respectively.}
\label{fig2}
\end{figure}

Figure~\ref{fig2} displays an interpolation between a healthy signal and a complete left bundle branch block (CLBBB) signal as observed in lead V1. The labels assigned to the signals are based on the occurrence of training samples for healthy signals, which are labeled as NORM and SR, and for CLBBB signals, which are labeled as SR and CLBBB. In an electrocardiogram (ECG), various main features are observed that are more representative of the CLBBB disease. Firstly, as the disease increases, the QRS complex widens, which is clearly visible in the 75\% and 100\% CLBBB signals. Secondly, on all signals, the RSR feature that is characteristic of CLBBB becomes larger as the disease progresses. Lastly, an upward trend on the T-wave can also be observed as the disease becomes more severe.

\subsubsection*{Left ventricular hypertrophy}
\begin{figure}[!ht]
\centerline{\includegraphics[width=\columnwidth]{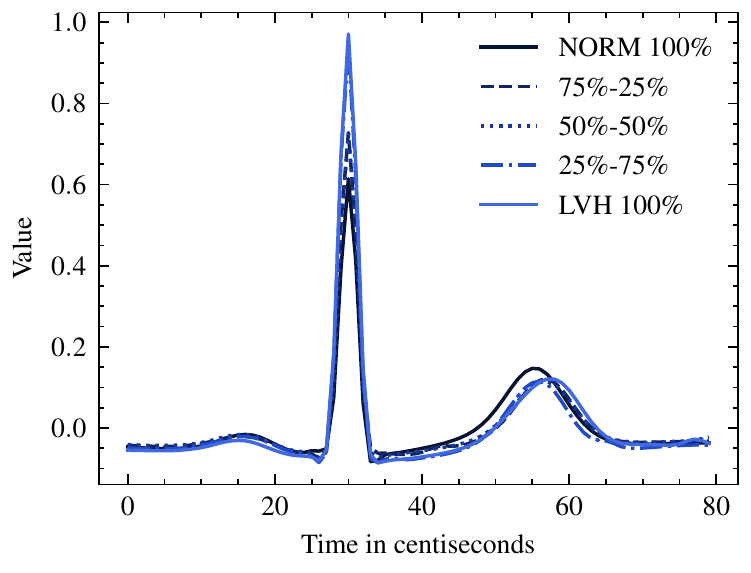}} 
\caption{\SSSDECG{} interpolation between a healthy (NORM+SR) and a left ventricular hypertrophy signal (LVH+SR+VCLVH) in lead V5, where 5 signals of 100\% and 0\%, 75\%, and 25\%, 50\% and 50\%, 25\% and 75\%, and 0\% and 100\% of healthy and left ventricular hypertrophy, respectively.}
\label{fig3}
\end{figure}

Figure~\ref{fig3} displays an interpolation between a healthy signal and a signal with left ventricular hypertrophy (LVH), as observed from lead V5. The labels used for healthy samples are NORM, SR, and for LVH samples, the labels used are SR, LVH and voltage criteria (QRS) for left ventricular hypertrophy (VCLVH), based on training sample occurrence. An important ECG feature of LVH can be observed, which is the enlargement of the R-peak in one of the left-side leads (V5). As the disease progresses, the R-peak also increases in size, and there is also a downward trend in the T-wave as the disease worsens.

In summary, these case interpolation studies, conducted across three different domains and aligned with domain knowledge on the considered conditions, suggest that \SSSDECG{} has gained a significant understanding of how different label combinations relate to specific signal features.

\subsection{Expert evaluation}

As final component of our evaluation of the sample quality, we presented the generated samples to an expert clinical and interventional cardiologist for qualitative assessment.

\subsubsection*{Generative diagnosis on normal samples}

To perform this task, we provided the expert with four 10-second 12-lead ECGs, including one real ECG and one sample from each of the generative models. These four complete ECG recordings offer a detailed visual representation of the generated samples. The expert was then asked to distinguish between real and synthetic signals.

\begin{figure}[!ht]
\centerline{\includegraphics[width=3.5in, height=3in]{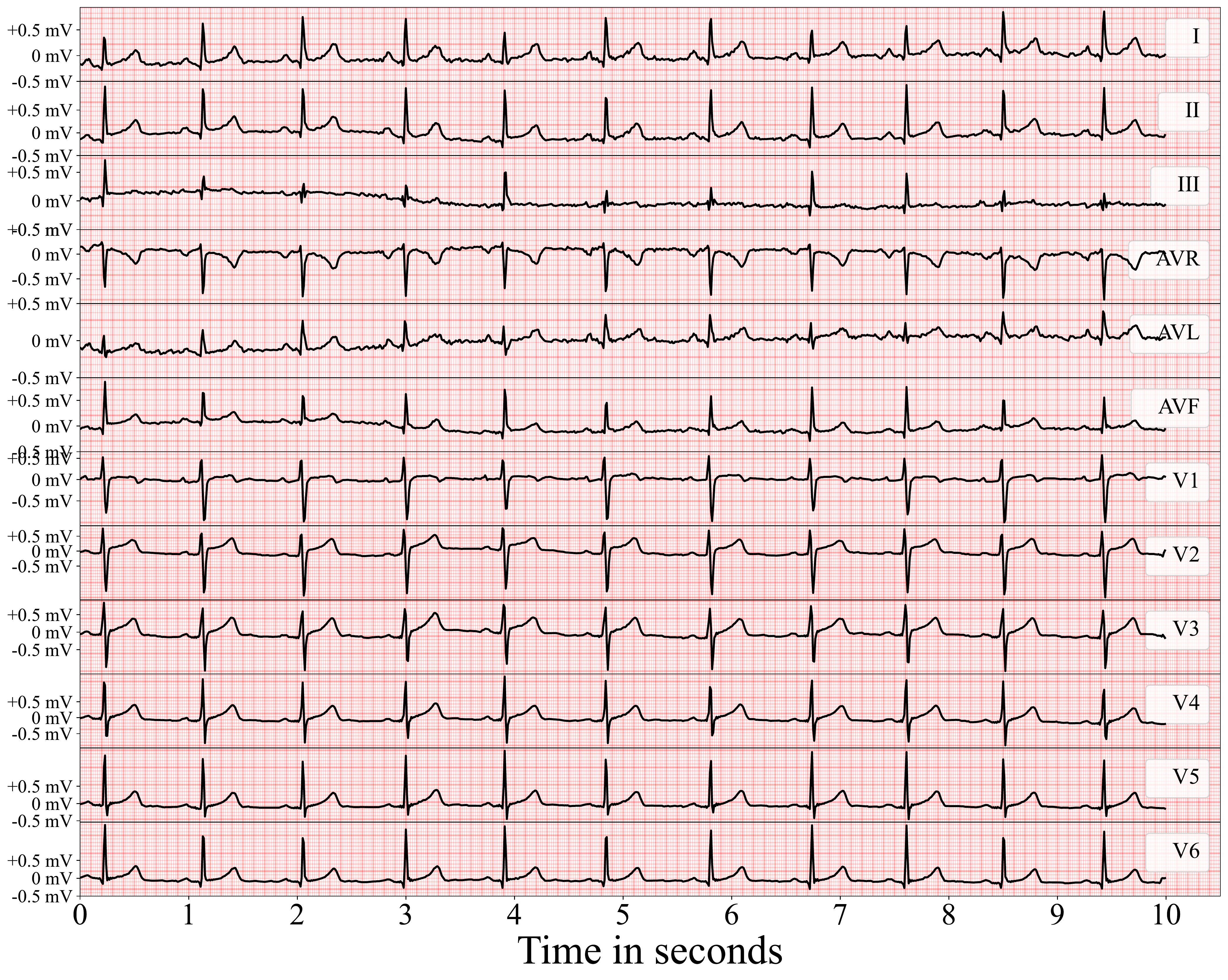}}  
\caption{Real (PTB-XL) NORM sample}
\label{fig10}
\end{figure}

\begin{figure}[!ht]
\centerline{\includegraphics[width=3.5in, height=3in]{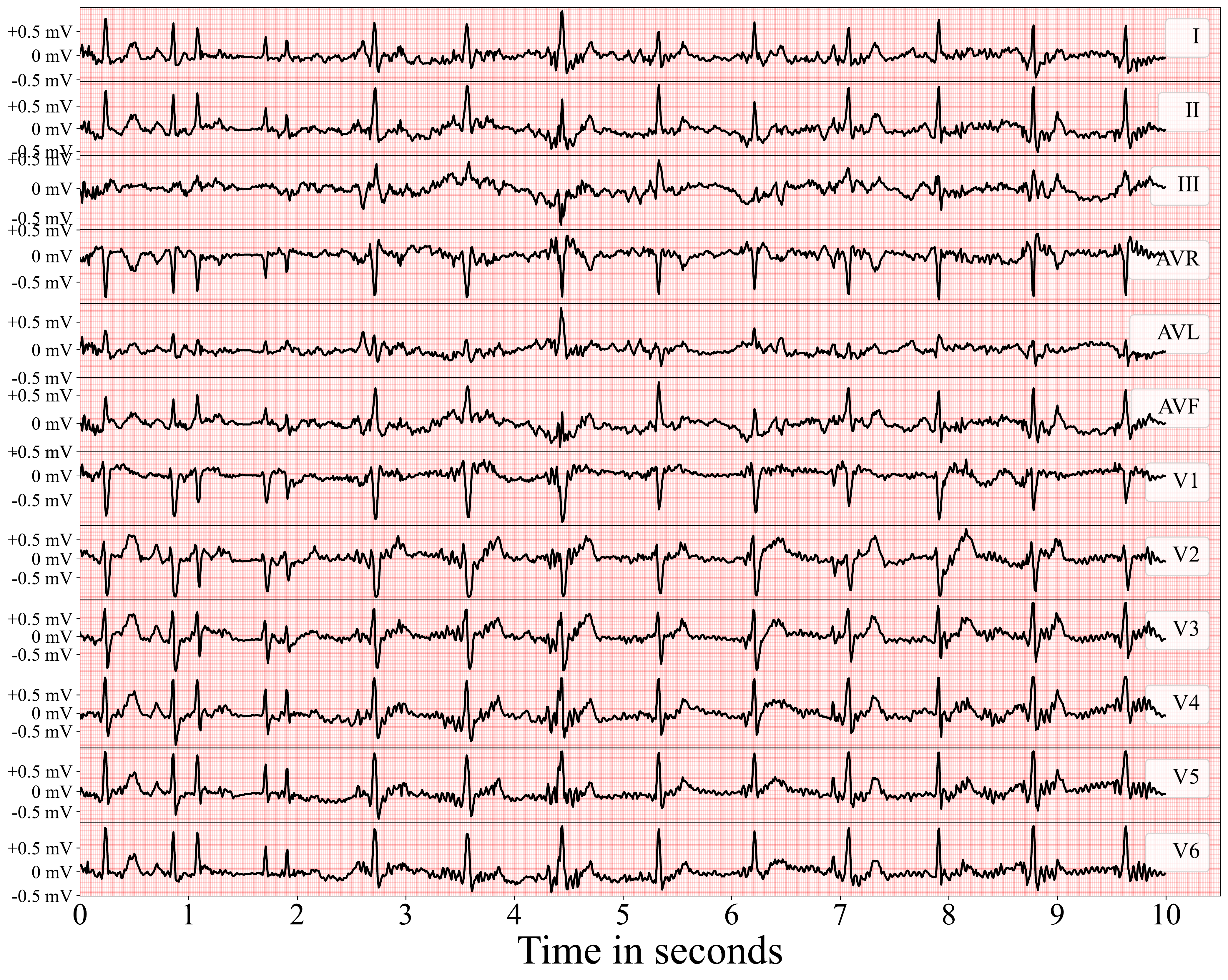}} 
\caption{Synthetic WaveGAN* NORM sample}
\label{fig11}
\end{figure}

\begin{figure}[!ht]
\centerline{\includegraphics[width=3.5in, height=3in]{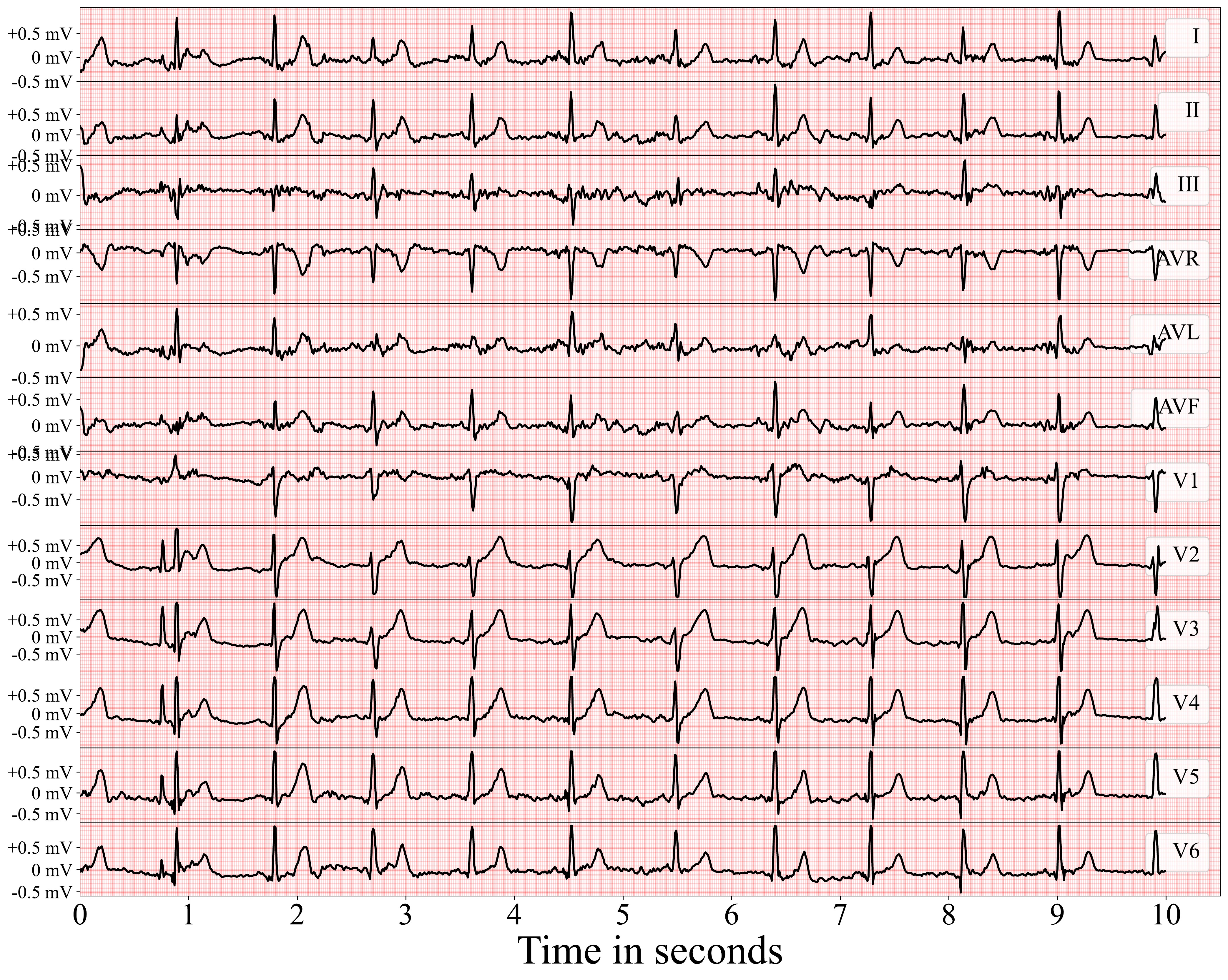}}
\caption{Synthetic Pulse2Pulse NORM sample}
\label{fig12}
\end{figure}

\begin{figure}[!ht]
\centerline{\includegraphics[width=3.5in, height=3in]{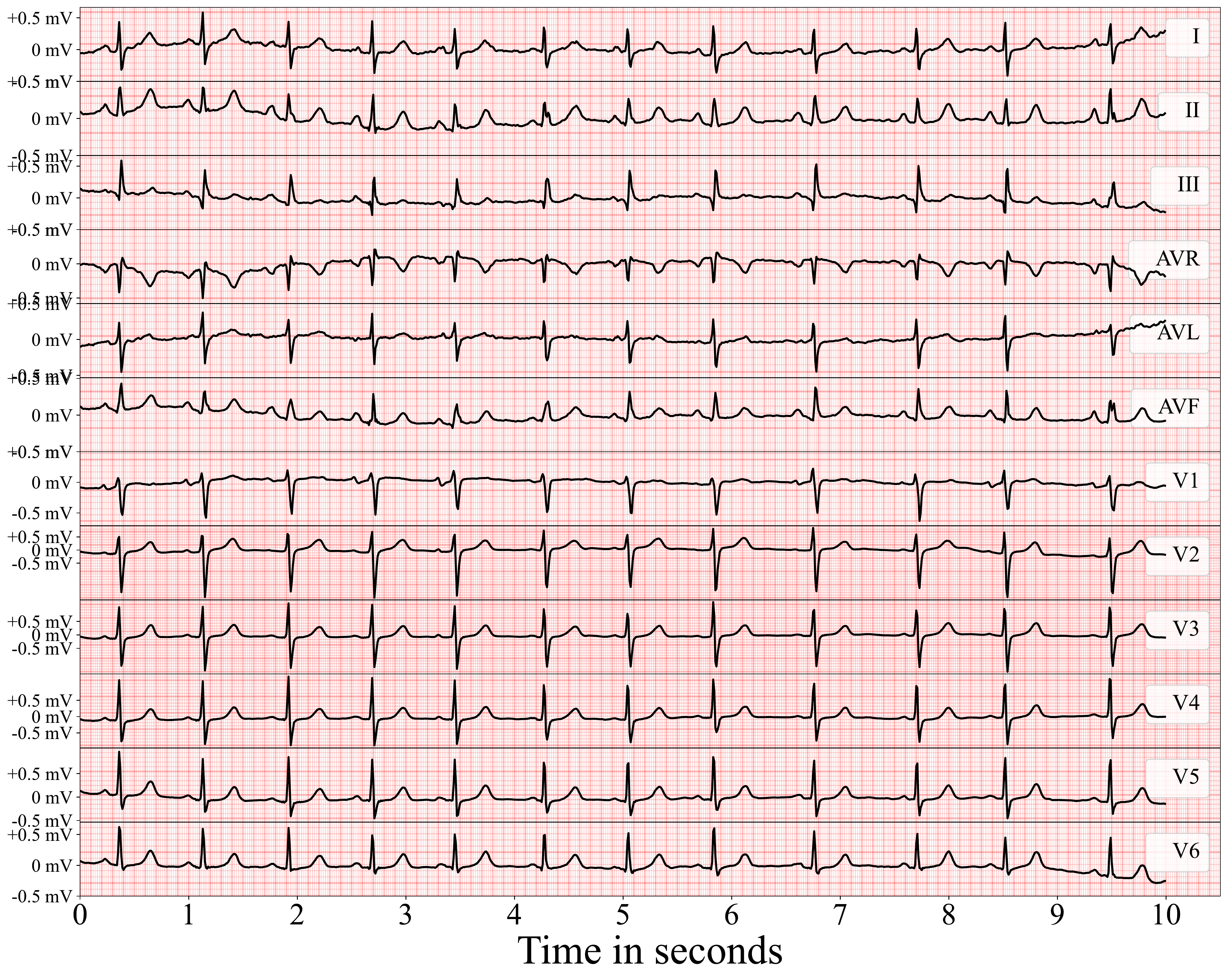}} 
\caption{Synthetic \SSSDECG{} NORM sample}
\label{fig13}
\end{figure}

Figure \ref{fig10} depicts a 10-second real NORM signal that was evaluated by the medical expert. The diagnosis was a synthetic sample because of changes in voltage in leads I, III, and AVF, and a slow first vector (R-wave) progression in the precordial leads V1-V6. Figure \ref{fig11} shows a 10-second WaveGAN* NORM signal that was also evaluated by a medical expert, and the diagnosis was a synthetic sample due to the absence of a sinus signal, unclear P-waves, high interference, no symmetry on RR intervals, and changes in voltage on the same trace. Figure \ref{fig12} depicts a 10-second Pulse2Pulse NORM signal that was also evaluated by a medical expert, and the diagnosis was a synthetic sample because of the variability in P-waves, no symmetry on RR intervals, high interference, and morphological changes in the same trace within the same lead. Lastly, Figure \ref{fig13} depicts a 10-second \SSSDECG{} NORM signal that was diagnosed as a real sample by the medical expert. The diagnosis was due to the well-defined P-wave and RR interval, and although there was a bit of tachycardia, the signal contained clear isodiphasic patterns, particularly in the I and AVF leads.

\subsubsection*{Clinical Turing test}

In this section, we present the results of a Turing test conducted with a medical professional previously mentioned. The task consisted of presenting 22 pairs of 12-leads ECGs (a total of 44), where one is synthetic (a total of 22) and the other is real (also a total of 22). The cardiologist was asked to perform two tasks. The first task was to decide if the sample was consistent with the provided set of ECG statements, which was either used to select a corresponding normal sample from PTB-XL or used to generate the synthetic sample. The second task was to determine if any of the ECGs in the given pair appeared synthetic. In Appendix~\ref{app:turing}, we provide a detailed breakdown of the various combinations involved, which correspond to the most frequent label combinations in PTB-XL, along with specific results for each pair. However, in this section, we will only report simple descriptive statistics. The original samples will be made available as part of the code repository\cite{coderepo}.

In the first part of the assignment, which involved evaluating diagnosis samples, a total of 44 samples were assessed. Among these, the cardiologist incorrectly diagnosed 19 (43.18\%) and correctly diagnosed 25 (56.81\%). Specifically, out of the 22 synthetic samples, 7 (31.81\%) were diagnosed incorrectly and 15 (68.18\%) were diagnosed correctly. Similarly, out of the 22 real samples, the cardiologist diagnosed 12 (54.54\%) incorrectly and 10 (45.45\%) correctly. The higher confirmation rate of 68\% for the synthetic samples, compared to 45\% for the real samples, indicates that the synthetic ECGs are very accurate and match their designated labels very well.

In the second part of the assignment, which involved identifying synthetic samples, a total of 44 samples were considered. Among these, the cardiologist identified 10 as synthetic, which represents 22.72\%, and 34 as real, which represents 77.27\%. Out of the 22 synthetic samples, the cardiologist identified 5 as synthetic (22.72\%) and 17 as real (77.27\%). Similarly, out of the 22 real samples, the cardiologist identified 5 as synthetic (22.72\%) and 17 as real (77.27\%). The fact that 5 samples were identified as synthetic in both cases confirms that it is difficult for a medical expert to distinguish between real and synthetic samples. It is worth emphasizing that this conclusion applies to a diverse set of 22 different medical conditions, which underscores the high quality of the synthetic samples generated.

\section{Conclusion}

In our study, we proposed a novel approach called \SSSDECG{} for generating electrocardiogram (ECG) data using diffusion-based techniques. We conditioned our model on a diverse set of 71 ECG statements in a multi-label setting, which is a highly complex task. The generated samples excelled in different context from qualitative over quantitative evaluation to conditional label interpolation and a human expert evaluation, clearly outperforming the two GAN-based competitors also proposed in this work. However, we observed that the generated \SSSDECG{} samples were not entirely capable of compensating for real samples when training a classifier on them. We believe that bridging this gap would be a significant achievement and a measurable sign of progress in the field in the near future. To encourage further research in this field, we are releasing the source code used in our investigations\cite{coderepo}, as well as the trained models and a synthetic copy of the PTB-XL dataset generated by \SSSDECG{} \cite{datarepo}.

\appendices

\section{PTB-XL dataset}
\label{app:ptbxl}

Table~\ref{tab: ptb-xl details} provides detailed information on the PTB-XL dataset \cite{Wagner:2020PTBXL,Wagner2020:ptbxlphysionet,Goldberger2020:physionet}, which is a collection of 21,837 clinical 12-lead electrocardiograms (ECGs) taken from 18,885 patients. Each ECG recording lasts for 10 seconds, and all signals were collected at a sampling rate of 100 Hz, i.e., corresponding 1000 time steps per sample. During the experiments, a set of 71 possible labels was used in a multi-label setting to provide conditional information.

\begin{table}[ht]
\caption{PTB-XL dataset details}
\centering
\setlength{\tabcolsep}{3pt}
\begin{tabular}{|p{75pt}|p{50pt}|}
\hline
  Description & Value \\ \hline\hline
  Train set size & 17,441\\
  Validation set size & 2,193\\
  Test set size & 2,203\\
  Sample length & 1000 \\ 
  Sample feature & 12 \\ 
  Labels &  71 \\ 
  \hline
\end{tabular}
\label{tab: ptb-xl details}
\end{table}

\section{XResNet1d50 details}
\label{app:xresnet}

Table~\ref{tab: resnet50} presents the XResNet1d(50) architecture details proposed in \cite{deep_ptbxl}, along with the corresponding training hyperparameters. The architecture comprises four blocks with 3, 4, 6, and 3 layers, respectively, of one-dimensional convolutions with an expansion of four and strides of 1. The training procedure employed the Adam optimizer with a learning rate and weight decay of \num{1e-3} for 100 epochs and a batch size of 64. Unlike the generated ECG data, the classifier was trained on cropped samples of 250 time steps each. During inference, the mean output probabilities were computed across seven different crops obtained by shifting the window through the signal using a stride of 125 time steps. The mean of the output probabilities was then used as the prediction for the entire sample.

\begin{table}[ht]
\caption{XResNet1d50 hyperparameters}
\centering
\setlength{\tabcolsep}{3pt}
\begin{tabular}{|p{75pt}|p{50pt}|}
\hline
  hyperparameter & Value \\ \hline\hline
  Block of layers & 4 \\
  Layers in each block & [3,4,6,3] \\ 
  Expansion & 4 \\
  Stride & 1 \\
  Optimizer & Adam \\ 
  Learning rate & \num{1e-3} \\ 
  Weight decay & \num{1e-3} \\ 
  Batch size & 64 \\ 
  Epochs & 100 \\ 
  \hline
\end{tabular}
\label{tab: resnet50}
\end{table}

\section{\SSSDECG{} details}
\label{app:sssdecg}

Table~\ref{tab: sssds4 hyperparameters} provides detailed information on the architecture of the \SSSDECG{} diffusion model, which consists of a network of 36 stacked residual layers with 256 residual and skip channels. For a comprehensive discussion of the model architecture, we refer the reader to \cite{lopezalcaraz2022diffusionbased}. The \SSSDECG{} model uses a swish activation function over the second and third levels and has a three-level diffusion embedding in 128, 256, and 256 dimensions. To compute the initial S4 diffusion, we constructed a convolutional layer after the diffusion embedding to double the input's residual channel dimension. We then employed a second S4 layer that increased the conditional information and its inclusion in the input. The output was routed through a gated-tanh non-linearity, and a convolutional layer was used to project residual channels back to the channel dimensionality. Regarding the hyperparameters, we used 200 time steps on a linear schedule for diffusion setup, with a beta value ranging from 0.0001 to 0.02. We employed Adam as an optimizer with a learning rate of $2\cdot 10^{-4}$. To learn the temporal dependencies of series in both directions for the S4 model, we used a single bidirectional S4 layer. Based on prior research \cite{Gu2021EfficientlyML}, we used layer normalization and an internal state dimensionality of $N=64$.

\begin{table}[ht]
\caption{\SSSDECG{} hyperparameters}
\centering
\setlength{\tabcolsep}{3pt}
\begin{tabular}{|p{120pt}|p{50pt}|}
 \hline
  Hyperparameter & Value \\  \hline\hline
  Residual layers & 36 \\ 
  Residual channels & 256 \\
  Skip channels & 256 \\ 
  Diffusion embedding dim. 1 & 128 \\ 
  Diffusion embedding dim. 2 & 512 \\ 
  Diffusion embedding dim. 3 & 512 \\ 
  Schedule  & Linear \\ 
  Diffusion steps $T$ & 200 \\ 
  $B_0$ & 0.0001 \\ 
  $B_1$ & 0.02 \\ 
  Optimizer & Adam \\ 
  Loss function & MSE  \\
  Learning rate & \num{2e-4} \\ 
  Batch size &   4 \\
  \hline
\end{tabular}
\label{tab: sssds4 hyperparameters}
\end{table}

\section{Baseline details}
\label{app:baselines}

\subsection{WaveGAN* details}

Table~\ref{tab: wavegan} displays the architecture and training hyperparameters utilized in the implementation of WaveGAN*. The generator takes in a one-dimensional vector of 1000 data points sampled from a uniform distribution and passes it through five deconvolution blocks to generate an output signal consisting of 1000 time steps and 8 ECG leads. Each deconvolution block is composed of four layers: an upsampling layer, a padding layer, a 1D-convolution layer, and a ReLU activation function, in that order. The discriminator and generator models have a size of 50 each. In the conditional setting, we incorporated a conditional batch normalization at every convolution in the generator. This batch normalization takes in a batch of labels to condition, which is passed through an embedding layer with an embedding dimension size twice that of the output channel dimensions, and then added to the convolution output.

\begin{table}[ht]
\caption{WaveGAN* hyperparameters}
\centering
\setlength{\tabcolsep}{3pt}
\begin{tabular}{|p{125pt}|p{30pt}|}
\hline
Hyperparameter    &   Value   \\ \hline\hline
Generator model size & 50 \\ 
Generator deconvolutional blocks  & 5 \\ 
Generator latent dimensions & 1000 \\ 
Discriminator model size & 50 \\ 
Discriminator convolutional blocks  & 6 \\ 
Optimizer & Adam \\ 
Loss function & MSE  \\
learning rate &     0.0001     \\ 
Training epochs       &     3000   \\
Batch size   &  32 \\ 
\hline
\end{tabular}
\label{tab: wavegan}
\end{table}

\subsection{Pulse2Pulse details}

Table~\ref{tab: pulse2pulse} presents the architecture and training hyperparameters utilized in the Pulse2Pulse implementation. Pulse2Pulse is an architecture that is akin to a U-Net, but differs in its use of one-dimensional convolutional layers for generating electrocardiogram (ECG) signals. The generator takes an input consisting of 1000 time steps and 8 channels sampled from a uniform distribution. This input undergoes five downsampling blocks and five upsampling blocks, where the upsampling technique used is similar to that of WaveGAN*. The downsampling process involves a one-dimensional convolution and a Leaky ReLU activation function. For the conditional setting, we added a conditional batch normalization to every convolution in the generator. This batch normalization takes in a batch of labels to condition the output, which is then passed through a word embedding with an embedding dimension size of twice the output channel dimensions. Finally, the result of the word embedding is added to the convolution output.

\begin{table}[ht]
\caption{Pulse2Pulse hyperparameters}
\centering
\setlength{\tabcolsep}{3pt}
\begin{tabular}{|p{125pt}|p{30pt}|}
\hline
Hyperparameter    &   Value   \\ \hline\hline
Generator model size & 50 \\ 
Generator deconvolutional blocks  & 5 \\ 
Discriminator model size & 50 \\ 
Discriminator convolutional blocks  & 6 \\ 
Optimizer & Adam \\ 
Loss function & MSE  \\
learning rate &     0.0001     \\ 
Training epochs       &     3000   \\ 
Batch size    &   32 \\
\hline
\end{tabular}
\label{tab: pulse2pulse}
\end{table}

\subsection{Other baselines}
\label{app:unrelieable_baselines}

In this final appendix, we would like to comment on several baseline models that we have tested for our use case. Specifically, we were unable to train functional generative models using TTS-GAN, TimeGAN, and cVAE\_ECG.

The authors presented a GAN model called TTS-GAN \cite{10.1007/978-3-031-09342-5_13}, as an unconditional generative model for time series data. They conducted various experiments to test the model, including one where they generated electrocardiogram (ECG) data. 
We implemented TTS-GAN with varying depths of layers for the generator and discriminator, including 4, 6, 12, and 24 layers. Additionally, we used different latent dimensions for the generator, such as 128, 256, and 1000. Since our sequence length was 1000, we mainly used two settings. The first setting had a patch size of 200 and an embedding dimension of 5, while the second setting had values of 100 and 10, respectively. The training process consisted of 200 epochs with a batch size of 4 and a learning rate of 0.0001 and 0.0003 for the generator and discriminator, respectively.
 
In a similar vein, TimeGAN \cite{NEURIPS2019_c9efe5f2} is a well-known GAN model used for generating unconditional time series data. Its creators conducted various experiments to test its efficacy, albeit mostly on short signal lengths. We attempted to train the TimeGAN model using the authors' recommended default hyperparameters, including a 3-layer GRU module with 24 hidden dimensions, a batch size of 4, and 2000 training iterations. However, we were unable to observe any meaningful generated samples.

Variational autoencoders (VAEs) are a popular framework for generative modeling. cVAE\_ECG \cite{9761431} is a conditional generative model specifically designed for ECG data. However, it differs from our approach in that it requires beat segmentation as a preprocessing step, rather than generating continuous signals . Additionally, the label set space of their implemented dataset is relatively small. In our work, we used a learning rate of 0.001, a batch size of 4, and various convolutional filter dimensions, including 3, 5, 8, and 10, as well as convolutional kernels at length dimensions of 10, 50, 100, 500, and 1000. We also experimented with latent spaces of 10, 20, 50, 100, and 1000, with a conditional dimensionality of 71 given the dataset labels. However, despite these efforts, we were unable to produce reasonable reconstructions using the given model architecture.

\section{Clinical Turing Test: Detailed results breakdown}
\label{app:turing}

Table~\ref{tab: turing test diagnosis evaluation} presents information about the initial phase of the Turing test, which involved evaluating diagnostic predictions. The predictions for sample A and sample B were represented by A(P) and B(P), respectively. The objective was to determine whether the given set of labels accurately matched the represented sample. True was used to denote correct diagnoses, while false indicated incorrect diagnoses.

\begin{table}[ht]
\caption{Turing test diagnosis evaluation}
\centering
\setlength{\tabcolsep}{3pt}
\begin{tabular}{|p{15pt}|p{85pt}|p{25pt}|p{25pt}|}
\hline
Set    &   Labels  & A(P) & B(P) \\ \hline\hline
1 & NORM, SR    & True & False \\
2  & NDT, SR   & False & True \\
3 & ABQRS, IMI, SR    & True & False  \\
4 & NORM, SARRH    & True & True  \\
5 & LAFB SR    & False & True   \\
6 & NORM SBRAD   & True & True  \\
7 & PACE    & True & True  \\
8 & CLBBB, SR     & True & False \\
9 & LVH, SR, VCLVH    & False & False  \\
10 & NORM   & True & False \\
11 & IRBBB, SR   & True & False \\
12 & ABQRS, NORM, SR   & True & True \\
13 & IRBBB, NORM, SR    & True & False  \\
14 & NORM, STACH    & False & True  \\
15 & IMI, SR   & False & True  \\
16 & ISC\_, LVH, SR   & True & True \\
17 & ABQRS, ASMI, SR   & True & False \\
18 & NST\_, SR   & False & True \\ 
19 & NDT, NT\_, SR   & True & False \\
20 & ABQRS, ASMI, IMI, SR   & False & True \\
21 & LVH, SR    & False & False \\
22 & AFIB, NST\_   & True & False \\
\hline
\end{tabular}
\label{tab: turing test diagnosis evaluation}
\end{table}

Table~\ref{tab: turing test synthetic evaluation} presents details regarding the synthetic evaluation that was conducted as the second part of the Turing test. In the table, A(T), B(T), A(P), and B(P) denote the true label for sample A or B, as well as the predicted labels for sample A or B, respectively. The value ``true'' indicates a synthetic sample, while ``false'' indicates a real sample.

\begin{table}[ht]
\caption{Turing test synthetic evaluation}
\centering
\setlength{\tabcolsep}{3pt}
\begin{tabular}{|p{15pt}|p{85pt}|p{25pt}|p{25pt}|p{25pt}|p{25pt}|}
\hline
Set    &   Labels    & A(T) & B(T)          & A(P) & B(P)  \\ \hline\hline
1 & NORM, SR              & False & True    & False & True \\
2  & NDT, SR              & False & True    & False & False \\
3 & ABQRS, IMI, SR        & True & False    & False & True \\
4 & NORM, SARRH           & True & False    & False & False \\
5 & LAFB SR               & False & True    & True & False \\
6 & NORM SBRAD            & True & False    & False & True \\
7 & PACE                  & False & True    & False & True \\
8 & CLBBB, SR             & False & True    & False & True  \\
9 & LVH, SR, VCLVH        & False & True    & False & True \\
10 & NORM                 & True & False    & False & False  \\
11 & IRBBB, SR            & True & False    & False & True  \\
12 & ABQRS, NORM, SR      & True & False    & False & False  \\
13 & IRBBB, NORM, SR      & False & True    & False & False  \\
14 & NORM, STACH          & False & True    & False & False   \\
15 & IMI, SR              & True  & False   & False & False   \\
16 & ISC\_, LVH, SR       & False & True    & False & False  \\
17 & ABQRS, ASMI, SR      & True  & False   & False & False   \\
18 &NST\_, SR             & True  & False   & False & False   \\
19 & NDT, NT\_, SR        & True  & False   & False & False  \\
20 & ABQRS, ASMI, IMI, SR & False & True    & False & True   \\
21 & LVH, SR              & False & True    & False & False \\
22 & AFIB, NST\_          & True  & False   & False & True \\
\hline
\end{tabular}
\label{tab: turing test synthetic evaluation}
\end{table}

\section*{Acknowledgment}
The authors would like to extend their gratitude to Erick Davila Zaragoza for conducting a meticulous evaluation of the samples that were generated. Mr. Zaragoza is a certified clinical and interventional cardiologist accredited by the Mexican National Council of Cardiology as a general practitioner from the University of Guadalajara (UDG), and as a cardiologist from the National Autonomous University of Mexico (UNAM).

\bibliography{main}{}

\begin{thebibliography}{10}
\providecommand{\url}[1]{#1}
\csname url@samestyle\endcsname
\providecommand{\newblock}{\relax}
\providecommand{\bibinfo}[2]{#2}
\providecommand{\BIBentrySTDinterwordspacing}{\spaceskip=0pt\relax}
\providecommand{\BIBentryALTinterwordstretchfactor}{4}
\providecommand{\BIBentryALTinterwordspacing}{\spaceskip=\fontdimen2\font plus
\BIBentryALTinterwordstretchfactor\fontdimen3\font minus
  \fontdimen4\font\relax}
\providecommand{\BIBforeignlanguage}[2]{{%
\expandafter\ifx\csname l@#1\endcsname\relax
\typeout{** WARNING: IEEEtran.bst: No hyphenation pattern has been}%
\typeout{** loaded for the language `#1'. Using the pattern for}%
\typeout{** the default language instead.}%
\else
\language=\csname l@#1\endcsname
\fi
#2}}
\providecommand{\BIBdecl}{\relax}
\BIBdecl

\bibitem{gdpr}
M.~Tzanou, \emph{Health Data Privacy under the GDPR: Big Data Challenges and
  Regulatory Responses}, ser. Routledge Research in the Law of Emerging
  Technologies.\hskip 1em plus 0.5em minus 0.4em\relax Taylor \& Francis, 2020.

\bibitem{hippa}
J.~Sullivan and A.~B. A. H.~L. Section, \emph{HIPAA: A Practical Guide to the
  Privacy and Security of Health Data}, ser. Hein's ABA Archive Microfiche
  Collection.\hskip 1em plus 0.5em minus 0.4em\relax Health Law Section,
  American Bar Association, 2004.

\bibitem{ecloud}
A.~Abbas and S.~U. Khan, ``A review on the state-of-the-art privacy-preserving
  approaches in the e-health clouds,'' \emph{IEEE Journal of Biomedical and
  Health Informatics}, vol.~18, pp. 1431--1441, 2014.

\bibitem{ehealthrecordsystem}
J.~L. Fernández-Alemán, I.~C. Señor, P.~Ángel Oliver~Lozoya, and A.~Toval,
  ``Security and privacy in electronic health records: A systematic literature
  review,'' \emph{Journal of Biomedical Informatics}, vol.~46, no.~3, pp.
  541--562, 2013.

\bibitem{bigdatahealthcare}
M.~M. Farooqi, M.~A. Shah, A.~Wahid, A.~Akhunzada, F.~Khan, N.~ul~Amin, and
  I.~Ali, \emph{Big Data in Healthcare: A Survey}.\hskip 1em plus 0.5em minus
  0.4em\relax Cham: Springer International Publishing, 2019, pp. 143--152.

\bibitem{blockchainhealth}
R.~Kumar, W.~Wang, J.~Kumar, T.~Yang, A.~Khan, W.~Ali, and I.~Ali, ``An
  integration of blockchain and ai for secure data sharing and detection of ct
  images for the hospitals,'' \emph{Computerized Medical Imaging and Graphics},
  vol.~87, p. 101812, 2021.

\bibitem{federatedlearning}
J.~Xu, B.~S. Glicksberg, C.~Su, P.~Walker, J.~Bian, and F.~Wang, ``Federated
  learning for healthcare informatics,'' \emph{Journal of Healthcare
  Informatics Research}, vol.~5, no.~1, pp. 1--19, Nov. 2020.

\bibitem{yin2021see}
H.~Yin, A.~Mallya, A.~Vahdat, J.~M. Alvarez, J.~Kautz, and P.~Molchanov, ``See
  through gradients: Image batch recovery via gradinversion,'' in
  \emph{Proceedings of the IEEE/CVF Conference on Computer Vision and Pattern
  Recognition}, 2021, pp. 16\,337--16\,346.

\bibitem{MADLEYDOWD201963}
P.~Madley-Dowd, R.~Hughes, K.~Tilling, and J.~Heron, ``The proportion of
  missing data should not be used to guide decisions on multiple imputation,''
  \emph{Journal of Clinical Epidemiology}, vol. 110, pp. 63--73, 2019.

\bibitem{classificationwithimputedmissingvalues}
T.~Shadbahr, M.~Roberts, J.~Stanczuk, J.~Gilbey, P.~Teare, S.~Dittmer,
  M.~Thorpe, R.~V. Torne, E.~Sala, P.~Lio, M.~Patel, A.-C. Collaboration,
  J.~H.~F. Rudd, T.~Mirtti, A.~Rannikko, J.~A.~D. Aston, J.~Tang, and C.-B.
  Schönlieb, ``Classification of datasets with imputed missing values: does
  imputation quality matter?'' \emph{arXiv preprint 2206.08478}, 2022.

\bibitem{ganslowdensity}
A.~Brock, J.~Donahue, and K.~Simonyan, ``Large scale gan training for high
  fidelity natural image synthesis,'' \emph{arXiv preprint 1809.11096}, 2018.

\bibitem{NEURIPS2021_49ad23d1}
P.~Dhariwal and A.~Nichol, ``Diffusion models beat gans on image synthesis,''
  in \emph{Advances in Neural Information Processing Systems}, vol.~34, 2021,
  pp. 8780--8794.

\bibitem{autoregressivelowfidelity}
R.~Child, S.~Gray, A.~Radford, and I.~Sutskever, ``Generating long sequences
  with sparse transformers,'' \emph{arXiv preprint 1904.1050ß}, 2019.

\bibitem{ecg_from_gans}
A.~M. Delaney, E.~Brophy, and T.~E. Ward, ``Synthesis of realistic ecg using
  generative adversarial networks,'' \emph{arXiv preprint 1909.09150}, 2019.

\bibitem{conditionalgenerativeaudiodatasets}
M.~Seibold, A.~Hoch, M.~Farshad, N.~Navab, and P.~F{\"u}rnstahl, ``Conditional
  generative data augmentation for clinical audio datasets,'' in \emph{Medical
  Image Computing and Computer Assisted Intervention--MICCAI 2022: 25th
  International Conference, Singapore, September 18--22, 2022, Proceedings,
  Part VII}.\hskip 1em plus 0.5em minus 0.4em\relax Springer, 2022, pp.
  345--354.

\bibitem{deepfakedigitalpatology}
K.~Falahkheirkhah, S.~Tiwari, K.~Yeh, S.~Gupta, L.~Herrera-Hernandez, M.~R.
  McCarthy, R.~E. Jimenez, J.~C. Cheville, and R.~Bhargava, ``Deepfake
  histologic images for enhancing digital pathology,'' \emph{Laboratory
  Investigation}, vol. 103, no.~1, p. 100006, 2023.

\bibitem{TOPOL2021785}
E.~J. Topol, ``What's lurking in your electrocardiogram?'' \emph{The Lancet},
  vol. 397, no. 10276, p. 785, 2021.

\bibitem{Wagner:2020PTBXL}
P.~Wagner, N.~Strodthoff, R.-D. Bousseljot, D.~Kreiseler, F.~I. Lunze,
  W.~Samek, and T.~Schaeffter, ``{PTB}-{XL}, a large publicly available
  electrocardiography dataset,'' \emph{Scientific Data}, vol.~7, no.~1, p. 154,
  2020.

\bibitem{Wagner2020:ptbxlphysionet}
P.~Wagner, N.~Strodthoff, R.-D. Bousseljot, W.~Samek, and T.~Schaeffter,
  ``{PTB-XL, a large publicly available electrocardiography dataset},'' 2020.

\bibitem{Goldberger2020:physionet}
A.~L. Goldberger, L.~A.~N. Amaral, L.~Glass, J.~M. Hausdorff, P.~C. Ivanov,
  R.~G. Mark, J.~E. Mietus, G.~B. Moody, C.-K. Peng, and H.~E. Stanley,
  ``{PhysioBank, PhysioToolkit, and PhysioNet},'' \emph{Circulation}, vol. 101,
  no.~23, pp. e215--e220, 2000.

\bibitem{Mehari:2022S4}
T.~Mehari and N.~Strodthoff, ``{Advancing the State-of-the-Art for ECG Analysis
  through Structured State Space Models},'' in \emph{arXiv}, 2022, extended
  abstract.

\bibitem{deep_ptbxl}
N.~Strodthoff, P.~Wagner, T.~Schaeffter, and W.~Samek, ``Deep learning for ecg
  analysis: Benchmarks and insights from ptb-xl,'' \emph{IEEE Journal of
  Biomedical and Health Informatics}, vol.~25, no.~5, pp. 1519--1528, 2021.

\bibitem{he2019bag}
T.~He, Z.~Zhang, H.~Zhang, Z.~Zhang, J.~Xie, and M.~Li, ``Bag of tricks for
  image classification with convolutional neural networks,'' in
  \emph{Proceedings of the IEEE/CVF Conference on Computer Vision and Pattern
  Recognition}, 2019, pp. 558--567.

\bibitem{pmlr-v37-sohl-dickstein15}
J.~Sohl-Dickstein, E.~Weiss, N.~Maheswaranathan, and S.~Ganguli, ``Deep
  unsupervised learning using nonequilibrium thermodynamics,'' in
  \emph{Proceedings of the 32nd International Conference on Machine Learning},
  ser. Proceedings of Machine Learning Research, vol.~37, 07--09 Jul 2015, pp.
  2256--2265.

\bibitem{chen2020wavegrad}
N.~Chen, Y.~Zhang, H.~Zen, R.~J. Weiss, M.~Norouzi, and W.~Chan, ``Wavegrad:
  Estimating gradients for waveform generation,'' in \emph{International
  Conference on Learning Representations}, 2020.

\bibitem{DBLP:conf/iclr/KongPHZC21}
Z.~Kong, W.~Ping, J.~Huang, K.~Zhao, and B.~Catanzaro, ``Diffwave: {A}
  versatile diffusion model for audio synthesis,'' in \emph{9th International
  Conference on Learning Representations, {ICLR} 2021}, 2021.

\bibitem{NEURIPS2020_4c5bcfec}
J.~Ho, A.~Jain, and P.~Abbeel, ``Denoising diffusion probabilistic models,'' in
  \emph{Advances in Neural Information Processing Systems}, vol.~33, 2020, pp.
  6840--6851.

\bibitem{Ho2022CascadedDM}
J.~Ho, C.~Saharia, W.~Chan, D.~Fleet, M.~Norouzi, and T.~Salimans, ``Cascaded
  diffusion models for high fidelity image generation,'' \emph{J. Mach. Learn.
  Res.}, vol.~23, pp. 47:1--47:33, 2022.

\bibitem{rombach2022high}
R.~Rombach, A.~Blattmann, D.~Lorenz, P.~Esser, and B.~Ommer, ``High-resolution
  image synthesis with latent diffusion models,'' in \emph{Proceedings of the
  IEEE/CVF Conference on Computer Vision and Pattern Recognition}, 2022, pp.
  10\,684--10\,695.

\bibitem{ho2022video}
J.~Ho, T.~Salimans, A.~Gritsenko, W.~Chan, M.~Norouzi, and D.~J. Fleet, ``Video
  diffusion models,'' \emph{arXiv preprint 2204.03458}, 2022.

\bibitem{Gu2021EfficientlyML}
A.~Gu, K.~Goel, and C.~R{\'e}, ``Efficiently modeling long sequences with
  structured state spaces,'' in \emph{International Conference on Learning
  Representations}, 2022.

\bibitem{NEURIPS2020_102f0bb6}
A.~Gu, T.~Dao, S.~Ermon, A.~Rudra, and C.~R\'{e}, ``Hippo: Recurrent memory
  with optimal polynomial projections,'' in \emph{Advances in Neural
  Information Processing Systems}, vol.~33, 2020, pp. 1474--1487.

\bibitem{9761431}
Y.~Sang, M.~Beetz, and V.~Grau, ``Generation of 12-lead electrocardiogram with
  subject-specific, image-derived characteristics using a conditional
  variational autoencoder,'' in \emph{2022 IEEE 19th International Symposium on
  Biomedical Imaging (ISBI)}, 2022, pp. 1--5.

\bibitem{RAHHAL2016340}
M.~A. Rahhal, Y.~Bazi, H.~AlHichri, N.~Alajlan, F.~Melgani, and R.~Yager,
  ``Deep learning approach for active classification of electrocardiogram
  signals,'' \emph{Information Sciences}, vol. 345, pp. 340--354, 2016.

\bibitem{deepfakeelectrocardiograms}
V.~Thambawita, J.~L. Isaksen, S.~A. Hicks, J.~Ghouse, G.~Ahlberg, A.~Linneberg,
  N.~Grarup, C.~Ellervik, M.~S. Olesen, T.~Hansen, C.~Graff, N.-H.
  Holstein-Rathlou, I.~Str{\"u}mke, H.~L. Hammer, M.~M. Maleckar, P.~Halvorsen,
  M.~A. Riegler, and J.~K. Kanters, ``\BIBforeignlanguage{en}{{DeepFake}
  electrocardiograms using generative adversarial networks are the beginning of
  the end for privacy issues in medicine},'' \emph{\BIBforeignlanguage{en}{Sci.
  Rep.}}, vol.~11, no.~1, p. 21896, Nov. 2021.

\bibitem{Zhu2019ElectrocardiogramGW}
F.~Zhu, F.~Ye, Y.~Fu, Q.~Liu, and B.~Shen, ``{Electrocardiogram generation with
  a bidirectional LSTM-CNN generative adversarial network},'' \emph{Scientific
  Reports}, vol.~9, 2019.

\bibitem{https://doi.org/10.48550/arxiv.1909.09150}
A.~M. Delaney, E.~Brophy, and T.~E. Ward, ``Synthesis of realistic ecg using
  generative adversarial networks,'' \emph{arXiv preprint 1909.09150}, 2019.

\bibitem{Golany_Lavee_Tejman_Yarden_Radinsky_2020}
T.~Golany, G.~Lavee, S.~Tejman~Yarden, and K.~Radinsky, ``Improving ecg
  classification using generative adversarial networks,'' \emph{Proceedings of
  the AAAI Conference on Artificial Intelligence}, vol.~34, no.~08, pp.
  13\,280--13\,285, Apr. 2020.

\bibitem{10.1007/978-3-031-09342-5_13}
X.~Li, V.~Metsis, H.~Wang, and A.~H.~H. Ngu, ``Tts-gan: A transformer-based
  time-series generative adversarial network,'' in \emph{Artificial
  Intelligence in Medicine}, M.~Michalowski, S.~S.~R. Abidi, and S.~Abidi,
  Eds.\hskip 1em plus 0.5em minus 0.4em\relax Cham: Springer International
  Publishing, 2022, pp. 133--143.

\bibitem{pmlr-v119-golany20a}
T.~Golany, K.~Radinsky, and D.~Freedman, ``{S}im{GAN}s: Simulator-based
  generative adversarial networks for {ECG} synthesis to improve deep {ECG}
  classification,'' in \emph{Proceedings of the 37th International Conference
  on Machine Learning}, ser. Proceedings of Machine Learning Research, H.~D.
  III and A.~Singh, Eds., vol. 119.\hskip 1em plus 0.5em minus 0.4em\relax
  PMLR, 13--18 Jul 2020, pp. 3597--3606.

\bibitem{Golany_Radinsky_2019}
T.~Golany and K.~Radinsky, ``Pgans: Personalized generative adversarial
  networks for ecg synthesis to improve patient-specific deep ecg
  classification,'' \emph{Proceedings of the AAAI Conference on Artificial
  Intelligence}, vol.~33, no.~01, pp. 557--564, Jul. 2019.

\bibitem{NEURIPS2019_c9efe5f2}
J.~Yoon, D.~Jarrett, and M.~van~der Schaar, ``Time-series generative
  adversarial networks,'' in \emph{Advances in Neural Information Processing
  Systems}, H.~Wallach, H.~Larochelle, A.~Beygelzimer, F.~d\textquotesingle
  Alch\'{e}-Buc, E.~Fox, and R.~Garnett, Eds., vol.~32.\hskip 1em plus 0.5em
  minus 0.4em\relax Curran Associates, Inc., 2019.

\bibitem{adib2023synthetic}
E.~Adib, A.~Fernandez, F.~Afghah, and J.~J. Prevost, ``Synthetic ecg signal
  generation using probabilistic diffusion models,'' 2023.

\bibitem{chung2023texttoecg}
H.~Chung, J.~Kim, J.-m. Kwon, K.-H. Jeon, M.~S. Lee, and E.~Choi,
  ``Text-to-ecg: 12-lead electrocardiogram synthesis conditioned on clinical
  text reports,'' in \emph{ICASSP 2023-2023 IEEE International Conference on
  Acoustics, Speech and Signal Processing (ICASSP)}.\hskip 1em plus 0.5em minus
  0.4em\relax IEEE, 2023, pp. 1--5.

\bibitem{lopezalcaraz2022diffusionbased}
J.~L. Alcaraz and N.~Strodthoff, ``Diffusion-based time series imputation and
  forecasting with structured state space models,'' \emph{Transactions on
  Machine Learning Research}, 2022.

\bibitem{coderepo}
J.~M.~L. Alcaraz and N.~Strodthoff, ``{SSSD-ECG public code repository},''
  \url{https://zenodo.org/account/settings/github/repository/AI4HealthUOL/SSSD-ECG},
  accessed: 2022-12-31.

\bibitem{goodfellow2020generative}
I.~Goodfellow, J.~Pouget-Abadie, M.~Mirza, B.~Xu, D.~Warde-Farley, S.~Ozair,
  A.~Courville, and Y.~Bengio, ``Generative adversarial networks,''
  \emph{Communications of the ACM}, vol.~63, no.~11, pp. 139--144, 2020.

\bibitem{de2017modulating}
H.~De~Vries, F.~Strub, J.~Mary, H.~Larochelle, O.~Pietquin, and A.~C.
  Courville, ``Modulating early visual processing by language,'' \emph{Advances
  in Neural Information Processing Systems}, vol.~30, 2017.

\bibitem{pmlr-v162-alaa22a}
A.~Alaa, B.~Van~Breugel, E.~S. Saveliev, and M.~van~der Schaar, ``How faithful
  is your synthetic data? {S}ample-level metrics for evaluating and auditing
  generative models,'' in \emph{Proceedings of the 39th International
  Conference on Machine Learning}, ser. Proceedings of Machine Learning
  Research, K.~Chaudhuri, S.~Jegelka, L.~Song, C.~Szepesvari, G.~Niu, and
  S.~Sabato, Eds., vol. 162.\hskip 1em plus 0.5em minus 0.4em\relax PMLR,
  17--23 Jul 2022, pp. 290--306.

\bibitem{datarepo}
J.~M.~L. Alcaraz and N.~Strodthoff, ``{SSSD-ECG data repository},''
  \url{https://figshare.com/s/43df16e4a50e4dd0a0c5}, accessed: 2022-01-19.

\end{thebibliography}
\bibliographystyle{IEEEtran}

\end{document}